# Heterogeneous peer effects of college roommates on academic performance

Yi Cao[1,2], Tao Zhou[1,2,*], Jian Gao[3,4,5,6,*]

[1] CompleX Lab, University of Electronic Science and Technology of China, Chengdu, China
[2] Big Data Research Center, University of Electronic Science and Technology of China, Chengdu, China
[3] Center for Science of Science and Innovation, Northwestern University, Evanston, IL, USA
[4] Kellogg School of Management, Northwestern University, Evanston, IL, USA
[5] Northwestern Institute on Complex Systems, Northwestern University, Evanston, IL, USA
[6] Faculty of Social Sciences, The University of Hong Kong, Hong Kong SAR, China
[*] Correspondence to zhutou@ustc.edu or jian.gao1@kellogg.northwestern.edu

## Abstract

Understanding how student peers influence learning outcomes is crucial for effective education management in complex social systems. The complexities of peer selection and evolving peer relationships, however, pose challenges for identifying peer effects using static observational data. Here we use both null-model and regression approaches to examine peer effects using longitudinal data from 5,272 undergraduates, where roommate assignments are plausibly random upon enrollment and roommate relationships persist until graduation. Specifically, we construct a roommate null model by randomly shuffling students among dorm rooms and introduce an assimilation metric to quantify similarities in roommate academic performance. We find significantly larger assimilation in actual data than in the roommate null model, suggesting roommate peer effects, whereby roommates have more similar performance than expected by chance alone. Moreover, assimilation exhibits an overall increasing trend over time, suggesting that peer effects become stronger the longer roommates live together. Our regression analysis further reveals the moderating role of peer heterogeneity. In particular, when roommates perform similarly, the positive relationship between a student's future performance and their roommates' average prior performance is more pronounced, and their ordinal rank in the dorm room has an independent effect. Our findings contribute to understanding the role of college roommates in influencing student academic performance.

# Introduction

Peer effects, or peer influence[1-5], have long been studied in the literature on social contagions[6-11] and education[12-18]. Understanding the influence of student peers on social behavior and learning outcomes is crucial for effective education management[18-22], as it can inform policy decisions on how to improve learning environments inside and outside the classroom[23-28]. Student peers can have both positive and negative effects, depending on their characteristics and behaviors[29,30]. For example, when surrounded by high-achieving peers, students may be motivated to improve their academic performance[31,32]. Meanwhile, some well-known examples of human behaviors adopted through social influence, such as smoking[33,34], substance abuse[35,36], and alcohol use[37-39], are often associated with negative student performance. Moreover, student peers may have indirect and lasting effects, for instance, on political ideology[40], persistence in STEM majors[41-45], occupational preferences[46], labor market outcomes[47-49], and earnings[50-53]. A thorough understanding of peer effects on learning outcomes can inform education management strategies, such as implementing behavioral interventions to mitigate the negative influence of disruptive peers[54,55]. Yet, using traditional methods and observational data to study peer effects causally is a challenge.

Dynamic educational and social environments make it difficult to separate peer influence from peer selection due to reverse causality, confounding factors, and complex mechanisms[1-3,56]. In particular, similarities in academic performance among student peers may be due to homophily (i.e., the selection of peers based on performance similarity) rather than the influence of peers[57-59]. Unlike open and evolving educational environments such as classrooms[23-26], dormitories in universities provide a close-knit living environment for students to interact and potentially learn from each other[60,61]. While dorm rooms may not be the primary learning place like classrooms and libraries, they offer a highly interpersonal and spillover environment for a small group of stable student peers. In contrast to Western universities in which freshman students usually have the flexibility to choose dormitories and suite-mates according to their lifestyle and personal preferences, most Chinese universities randomly assign students to dorm rooms[61-63]. There, a typical 4-person dorm room contains four beds and some public areas, providing a more interactive environment than a Western dorm suite containing four bedrooms (Supplementary Figure 1).

Research on student peer effects, on the one hand, has primarily relied on static observational data of campus behaviors and performance metrics[11,64]. This reliance stems from various factors, such as the high cost and impracticality of conducting large-scale field experiments in learning environments, the dynamic nature of peer relationships[65], and the scarcity of longitudinal data on learning outcomes[66-70]. The close-knit dormitory environment of Chinese universities, however, provides a unique opportunity to observe a stable group of student peers and track their academic performance over time[61,63]. On the other hand, while regression models are widely employed in studying peer effects within the social sciences, methodologies from other disciplines may help expand the functional form in which peer effects can be estimated[64]. Particularly, null models are well suited for studying non-trivial features of complex systems by providing an unbiased random structure of other features[71-73]. Null-model approaches have been applied to test causal effects in complex social systems[74-76]. For instance, in the social network literature, randomizations are used to study the impact of network interventions on social relationships[77]. Utilizing a null model to test whether roommates exhibit similar performance could offer a promising approach to identifying peer effects and quantifying their magnitude, facilitating comparisons across diverse datasets.

One advantage of canonical regression models is their capability to address the issue of inverse causality by utilizing longitudinal data and controlling for confounding factors[68,78]. For example, a student's future performance may be influenced not only by the average prior performance of roommates but also by their prior performance. Additionally, the composition of roommates may have independent effects[79]. Yet, it remains relatively less explored whether the heterogeneity in performance among roommates provides a ladder for the student to catch up with high-achieving roommates or hamper their motivation due to the inconsistent signal from roommates or the negative impact of disruptive roommates[29,30]. Moreover, dorm rooms provide an interactive yet local environment where a student's ordinal rank in the dorm room, conditional on academic performance, may independently affect learning outcomes[80,81]. Therefore, a more comprehensive understanding of the factors contributing to roommate peer effects may help inform education policy and student management strategies, such as designing interventions for dormitories that effectively leverage the influence of high-achieving peers in improving student performance.

In this study, we quantify roommate peer effects using both null models and regression approaches to analyze a longitudinal dataset of student accommodation and academic performance. Sourced from a public research-intensive university in China, our data covers 5,272 undergraduate students residing in 4-person dorm rooms following the random assignment of roommates (see Methods). The initialization is plausibly random since the roommate assignment takes into account neither students' academic performance before college admission nor their personal preferences, and there is no significant reassignment later (see Supplementary Information Section 1.2 for details). Here, we demonstrate the presence of roommate peer effects by showing that roommates with similar performance are more likely to be observed in the actual data than expected by chance alone. We then measure the size of roommate peer effects by developing an assimilation metric of academic performance and contrasting its value in the actual data with that in the roommate null model that we construct by randomly shuffling students among dorm rooms while retaining their controlled characteristics. Further, we use regression models to examine factors influencing roommate peer effects and explore the role of peer heterogeneity in moderating the effects.

## Results

**Tier combinations within a dorm room**

We start by studying the roommate composition of a typical 4-person dorm room in terms of their academic performance. For comparisons across student cohorts (i.e., those who were admitted by the university in the same year), majors, and semesters, we transform each student's grade point average (GPA) in a semester into the GPA percentile $R$ among students in the same cohort and major, where $R = 0$ and $R = 1$ correspond to the lowest and highest academic performance, respectively. We then divide students into equal-sized tiers based on their GPA percentiles, where those with better performance are in larger tiers. For instance, under the 4-tier classification, students with $R = 0.3$ (i.e., GPA is above 30% of students) and $R = 0.9$ (i.e., GPA is above 90% of students) are in Tier 2 and Tier 4, respectively. Accordingly, each dorm room has a tier combination without particular order. For example, 3444 (i.e., one student is in Tier 3, and the other three are in Tier 4) is identical to 4344 and 4434. Here we use the one in ascending order of tier numbers to delegate all identical ones. Under the 2-tier classification, there are five unique tier combinations (1111, 1112, 1122, 1222, and 2222). The numbers are 15 and 35 under 3-tier and 4-tier classifications, respectively (Fig. 1a; see Supplementary Information Section 2.1 for details).

Given a tier for classification, the probability $P_a$ of observing a combination in the actual data can be calculated by the fraction of dorm rooms with the combination. The actual probabilities $P_a$ of observing different combinations (i.e., the frequency of observations), however, shouldn't be directly compared. This is because their theoretical probabilities $P_t$ are not always the same even when the tier numbers of roommates are independent of each other, i.e., there is no roommate peer effect (see Supplementary Table 1 and Supplementary Information Section 2.1). To give a simple example: under the 2-tier classification, the theoretical probability $P_t$ of combination 1112 is $C_4^1 \left(\frac{1}{2}\right)^3 \left(\frac{1}{2}\right) = \frac{1}{4}$, which is four times as big as that of combination 1111, namely, $\left(\frac{1}{2}\right)^4 = \frac{1}{16}$. This leads to the difficulty of assessing, by the value of $P_a$, whether a combination is over-represented or under-represented in the actual data. To address this challenge, we calculate the relative ratio $\mathbb{E}$ for a combination by comparing the actual probability with its theoretical probability:

$$\mathbb{E} = \frac{P_a - P_t}{P_t}, \tag{1}$$

where $P_a$ and $P_t$ are the actual and theoretical probability of the same combination, respectively. A positive (negative) value of $\mathbb{E}$ suggests that the combination is more (less) likely to be observed in data than expected by chance alone (see Supplementary Information Section 2.2 for details).

We analyze the student accommodation and academic performance data under 2-tier, 3-tier, and 4-tier classifications and calculate the relative ratio $\mathbb{E}$ for each combination (Fig. 1a). We find that $\mathbb{E}$ of different combinations varies substantially and $\mathbb{E}$ of some combinations deviates significantly from 0 according to the results of statistical tests (see Methods and Supplementary Information Section 3.2 for details). For example, under the 2-tier classification, $\mathbb{E}$ of combinations 1111 and 2222 is significantly above 0 and $\mathbb{E}$ of combinations 1112 and 1122 is significantly below 0 ($P$ value < 0.001; see Supplementary Table 2 for the statistical testing results for each combination). More notably, we find that combinations with the same or nearby tier numbers (e.g., 1111 and 1112) tend to have larger $\mathbb{E}$ and those with distant tier numbers (e.g., 1122) have smaller $\mathbb{E}$, prompting us to study the relationship between a combination's tier heterogeneity and its $\mathbb{E}$. Specifically, we first calculate the relative difference $D$ in the tier numbers for each combination:

$$D = \frac{1}{6} \sum_{u \neq v} |l_u - l_v|, \quad 1 \leq u < v \leq 4, \tag{2}$$

where $l_u$ and $l_v$ is the tier number of roommates $u$ and $v$, respectively. A smaller $D$ indicates that roommates have closer tier numbers and thus a smaller difference in their academic performance. We then group combinations with the same $D$ and arrange them in ascending order of $D$. We find that combinations with positive and negative $\mathbb{E}$ are overall separated (Fig. 1b), where those with a smaller $D$ tend to have positive $\mathbb{E}$ (i.e., over-represented in the actual data) and those with a larger $D$ tend to have negative $\mathbb{E}$ (i.e., underrepresented in the actual data). Inspired by this observation, we calculate the mean value of $\mathbb{E}$ for each group with the same $D$, finding a negative relationship between $D$ and $\mathbb{E}$ (Fig. 1c). These results demonstrate that roommates tend to have more similar academic performance than random chance, suggesting the presence of roommate peer effects.

**Assimilation of roommate academic performance**

We generalize the tier combination analysis to the most granular tier for classification by directly dealing with the GPA percentile $R \in [0, 1]$ (hereafter GPA for short). Specifically, similar to calculating the relative difference $D$ in the tier combination for each dorm room, we develop an assimilation metric $A$ to quantify the extent to which the GPAs of roommates differ from each other. Formally, the assimilation metric $A$ for a 4-person dorm room is calculated by

$$A = 1 - \frac{1}{4}\sum_{u \neq v}|R_u - R_v|, 1 \leq u < v \leq 4, \quad (3)$$

where $R_u$ and $R_v$ are the GPAs of roommates $u$ and $v$, respectively. The assimilation $A$ of a dorm room is between 0 and 1 with a larger value indicating that roommates have more similar academic performance. If there is no roommate peer effect, each roommate's GPA should be independent and identically distributed (i.i.d.), and the theoretical assimilation $A$ of all dorm rooms has a mean value of 0.5 (see Supplementary Information Section 4.1 for detailed explanations).

Inspired by permutation tests, often referred to as the "quadratic assignment procedure" in social network studies[74,75], we perform a statistical hypothesis test to check whether the assimilation of dorm rooms in the actual data deviates significantly from its theoretical value. Specifically, we proxy theoretical assimilation via null-model assimilation that is calculated based on a roommate null model and compare it with actual assimilation. An appropriate null model of a complex system satisfies a collection of constraints and offers a baseline to examine whether displayed features of interest are truly non-trivial[71-73]. We start with the actual roommate configuration and randomly shuffle students between dorm rooms while preserving their compositions of cohort, gender, and major. By repeating this process, we construct a plausible roommate null model that consists of 1000 independent implementations (see Supplementary Information Section 3.1 for details). We find that the mean of actual assimilation (0.549) of all dorm rooms is 10.7% larger than that of null-model assimilation (0.496; Fig. 2a). A student's *t*-test confirms that the two assimilation distributions have significantly different means ($P$ value < 0.001; see Supplementary Information Section 4.2 for details). These results suggest that roommate assimilation in academic performance is greater than expected by chance, demonstrating significant roommate peer effects.

The extent to which the mean of actual assimilation is larger than that of null-model assimilation indicates the magnitude of roommate peer effects, allowing us to examine temporal trends over the five semesters. First, we find that roommate peer effects remain significant when measured using data from each semester (see Supplementary Information Section 4.1 for details). Second, we hypothesize that before the first semester (i.e., the first day of college), roommate peer effects should be 0 due to the plausible random assignment of roommates, where the actual assimilation should be close to the null-model assimilation. As roommates live together longer and establish stronger interactions with each other, the actual assimilation of roommate academic performance would become larger, and the magnitude of roommate peer effects would become bigger. To test this hypothesis, for each semester we calculate the percentage difference in the means of the actual assimilation and the null-model assimilation that is a proxy of roommate peer effects before the first semester (see Supplementary Information Section 4.1 for details). We find that the percentage difference exhibits an overall increasing trend over time (Fig. 2b), which supports the hypothesis that, as roommates live together longer, the magnitude of roommate peer effects on academic performance becomes larger. These results are robust when we use an alternative way to estimate the magnitude of roommate peer effects, where we calculate the share of dorm rooms with larger-than-null-model assimilation (see Supplementary Information Section 4.3 for details). Moreover, our further analysis shows that female and male students have similar assimilation, suggesting no significant gender differences (see Supplementary Information Section 4.4 for details).

**The effects of heterogeneous peers**

The increasing assimilation of roommates in their academic performance raises a question about how a student's future performance is impacted by their roommates' prior performance, especially when there is substantial peer heterogeneity in performance, e.g., there are both high-achieving and underachieving roommates. To answer this question, we employ regression models to perform a Granger causality type of statistical analysis. Specifically, we first examine the relationship between a student's post-GPA (GPA_Post; e.g., their own GPA in the second semester) and the average prior GPA of their roommates (RM_Avg; e.g., their roommate's average GPA in the first semester) by calculating pairwise correlations for all consecutive semesters and dorm rooms. We find that dorm rooms tend to occupy the diagonal of the "GPA_Post – RM_Avg" plane (Fig. 3a), suggesting that a student's post-GPA is positively associated with the average prior GPA of their roommates. We then use an ordinary least squares (OLS) model to study the relationship between GPA_Post and RM_Avg (see Methods for the empirical specification) and summarize the regression results in Table 1. We find that without controlling for the effects of other factors (see column (1) of Table 1), the average prior GPA of roommates has a significantly positive effect on a student's post-GPA (regression coefficient $b = 0.365$; $P$ value $< 0.001$; Fig. 3b).

Other factors may independently affect a student's post-GPA and confound its association with the average prior GPA of their roommates. Therefore, we add controls and fixed effects into the OLS model (see Methods). The regression results shown in Table 1 convey several findings. First, a student's prior GPA has the strongest effect on their post-GPA ($b = 0.801$, which is 16 times as large as $b = 0.050$ for roommate average prior GPA; see columns (2) of Table 1), suggesting a significant path dependence on academic achievement. Second, the positive effect of roommate average prior GPA on a student's post-GPA remains significant with controlling the student's prior GPA, gender, cohort, major, and semester ($P$ value < 0.01; see column (2) of Table 1 and Fig. 3c). Notably, female students perform better than male students on average (see Supplementary Information Section 4.4 for details). Third, the differences in roommate prior GPAs (RM_Diff) have no significant effect ($P$ value > 0.1; see columns (3) and (4) of Table 1), but it significantly moderates the relationship between roommate average prior GPA and post-GPA (see column (5) of Table 1 and Fig. 3d) such that their positive relationship is more pronounced (slope $b = 0.055$; 95% CI = [0.040, 0.070]) when RM_Diff is high (i.e., 1 s.d. above its mean) and less pronounced (slope $b = 0.028$; 95% CI = [−0.001, 0.057]) when RM_Diff is low (i.e., 1 s.d. below its mean; see Supplementary Information Section 5.1 for detailed results of a simple slope test). The result also shows that high post-GPA is associated with large differences in roommate prior GPA when roommate average prior GPA is low (see the red line on the lower left of Fig. 3d).

While the regression results suggest that roommate peer effects are significant, it is worth noting that the effect size appears to be modest. Specifically, a 100-point increase in roommate average prior GPA is associated with a 5-point increase in post-GPA ($b = 0.050$; see column (4) of Table 1). The effect is about 6% as large as the effect of a 100-point increase in prior GPA ($b = 0.801$), and it is about 10% of the average post-GPA. The magnitude is at a similar scale as reported by prior studies for various environments (e.g., dormitories and classrooms) and cultures (e.g., Western universities; see Supplementary Information Section 5.1 for details). To demonstrate its significance, we perform a falsification test by running the same OLS regression on the roommate null model, finding that the reported results are nontrivial (see Supplementary Information Section 5.3 for details). Together, these regression results suggest that a student's performance is impacted not only by the average performance of roommates but also by their heterogeneity.

**The effects of in-dorm ordinal rank**

Dorm rooms provide a highly interpersonal yet local environment, where competitive dynamics between roommates may affect their academic performance. Conditional on absolute academic performance, the ordinal rank of a student in their dorm room could have an independent effect on future achievement[80,81]. For instance, when a student's ordinal rank is consistently low across all semesters even if their absolute performance is high (e.g., the student has a GPA $R = 0.9$ and their roommates all have $R > 0.9$), they may still feel discouraged and less motivated, leading to fewer interactions with others and a potential decline in performance (see Supplementary Information Section 5.2 for explanations). This motivates us to study how a student's in-dorm ordinal rank (OR_InDorm, with 1 being the highest and 4 being the lowest according to their prior performance; i.e., the number of better-achieving roommates including themselves) affects their post-GPA. Specifically, we employ an OLS model that not only controls the student's prior GPA, their roommate average prior GPA and differences in prior GPAs, gender, and semester but also includes the fixed effects of cohort and major (see Methods for the empirical specification). We find that ordinal rank has a significantly positive effect on post-GPA ($P$ value < 0.05; see columns (1) and (2) of Table 2 and Fig. 4a), suggesting that the number of better-achieving roommates in the dorm room predicts a student's better academic performance in the future.

Through regression, we further examine whether the positive relationship between ordinal rank and post-GPA is moderated by other factors. We find that neither the interaction term of ordinal rank and own prior GPA nor the interaction term of ordinal rank and average roommate prior GPA is significant ($P$ value > 0.1; see columns (3) and (4) of Table 2). Yet, the interaction term of ordinal rank and differences in roommate prior GPA (RM_Diff) is significantly negative (*P* value < 0.05; see columns (5) of Table 2). Specifically, the effect of ordinal rank on post-GPA is more pronounced (slope *b* = 0.007; 95% CI = [0.002, 0.012]) when RM_Diff is low (Fig. 4b), while the effect is not significant (slope *b* = −0.000; 95% CI = [−0.007, 0.007]) when RM_Diff is high (see Supplementary Information Section 5.2 for detailed results of a simple slope test). The result also shows that high post-GPA is associated with large differences in roommate prior GPA when ordinal rank is low (see the red line on the lower left of Fig. 4b). Although the effect size is modest, our falsification test on the roommate null model demonstrates that the results are nontrivial and significant (see Supplementary Information Section 5.3 for details). Taken together, these results suggest that roommate peer effects tend to disproportionately benefit underachieving students with homogeneous roommates (i.e., those who have similar performance) and high-achieving students with heterogeneous peers (i.e., those who have widely varied performance).

**Discussion**

We quantified roommate peer effects on academic performance by applying both null-model and regression approaches to analyze a longitudinal dataset of student accommodation and academic performance, where roommate assignments are plausibly random upon enrollment and roommate relationships persist until graduation. We found evidence showing that roommates have a direct influence on a student's performance, with some heterogeneity in the variation among the roommates and the baseline achievement of the student. Specifically, by constructing a roommate null model and calculating an assimilation metric, we showed that roommates have more similar performance than expected by chance alone. Moreover, the average assimilation of roommate academic performance exhibits an overall increasing trend over time, suggesting that peer effects become stronger as roommates live together longer, get more familiar with each other, and establish stronger interactions that facilitate knowledge spillovers[61,65,82]. More specifically, the increase in assimilation is more pronounced in the third semester (Fig. 2b and Supplementary Figure 8), which is consistent with previous literature showing that peer effects are strong and persistent when friendships last over a year[79,83], and it appears to be disrupted in the fifth semester, which may be because senior students have a higher chance of taking different elective courses and have more outside activities that might decrease the interactions between roommates[84].

Our regression analysis further unpacks roommate peer effects, especially along the dimension of peer heterogeneity. We found that a student's future performance is not only strongly predicted by their prior performance, suggesting a significant path dependence in academic development[85-87], but also impacted by their roommates' prior performance. Also, the positive relationship between a student's future performance and the average prior performance of roommates is moderated by peer heterogeneity such that it is more pronounced when roommates are similar. In particular, when living with roommates who have on average low prior performance, a student benefits more if roommates are more different, suggesting the positive role of peer heterogeneity[88-90]. Moreover, ordinal rank in the dorm room has an independent effect since the number of better-achieving roommates is positively associated with future performance. Yet, peer heterogeneity moderates this relationship such that it is significant only when roommates are more similar. The magnitudes of peer effects assessed using regression may appear modest but they are significant and in line with the literature. Together, these results paint a rich picture of roommate peer effects and suggest that the effective strategy for improving a student's performance may depend on their position in a high-dimensional space of ordinal rank, peer average performance, and peer heterogeneity.

While our work helps better understand roommate peer effects, the results should be interpreted in light of the limitations of the data and analysis. First, the longitudinal data were limited to two cohorts of Chinese undergraduates in one university. The extent to which these findings can be generalized to other student populations, universities, and countries should be further investigated where relevant data on student accommodation and academic outcomes are available. Second, the roommate assignments were plausibly random according to the administrative procedures. While providing supporting evidence for this assumption (see Supplementary Information Section 1.2 for details), we lacked comprehensive data on student demographics, personal information, and pre-college academic performance to examine it directly. Third, the analysis relies on GPA percentiles normalized for each cohort and major, which allows for fair comparisons between disciplines but, at the same time, may lose more information in the data. A better normalization that preserves the distribution of GPAs, for example, would be an improvement. Fourth, factors outside of the dormitory environment may mediate the assimilation of roommates' academic performance such as orderliness, classroom interactions, social networks, behavior patterns, and common external factors[16,17,65]. Unraveling the mechanisms underlying roommate peer effects (e.g., peer pressure and student identity[91]) was beyond the reach of this study but is desirable as future work.

In summary, we demonstrate the peer effect of college roommates and assess its magnitude by employing canonical statistical methods to analyze new longitudinal data from a quasi-experiment. The university dorm room environment is ideal for identifying a group of frequently interacting and stable student peers whose learning outcomes can be easily tracked. The null model we use, which is essentially permutation tests[75,76], does not assume linear relationships between variables and is flexible enough to be applied to study peer effects in other complex social systems. Also, effect sizes assessed by the null model can facilitate comparisons between different datasets. The regression model allows us to address concerns about inverse causality and better understand peer effects. Particularly, the regression findings have potential policy implications for education and dormitory management. For example, by adjusting the composition of roommates such as reducing peer heterogeneity for students with on average high-achieving roommates, dorm rooms may be engineered, to some extent, to enhance the positive influence of roommates in improving students' academic performance. Furthermore, our findings suggest the benefits of exposure to student role models and learning from peers in everyday life in addition to teachers in classrooms only.

**Methods**

**Data.** Chinese universities provide on-campus dormitories for almost all undergraduates, allowing us to observe a large-scale longitudinal sample of student roommates and relate it to their academic performance. From a public university in China, we collected the accommodation and academic performance data of 5,272 undergraduates, who lived in identical 4-person dorm rooms in the same or nearby dorm building on campus. Different from a dorm suite that contains 4 bedrooms, a 4-person dorm room is a single bedroom with 4 beds, where each student occupies one bed and shares public areas with roommates (see Supplementary Figure 1 for an example layout). Per the university's student accommodation management regulations, newly admitted students were assigned to dorm rooms under the condition that those in the same administrative unit, major, or school live together as much as possible and there is no gender mix in dorm rooms or buildings. The process neither allowed students to choose roommates or rooms nor took into account their academic performance before admission, socioeconomic backgrounds, or personal preferences. Students were informed of their accommodation only when they moved in before the first semester. As a quasi-experiment, the administrative procedure resulted in a plausibly if not perfect, random assignment of roommates concerning their prior academic performance and personal information. Moreover, there was no significant individual selection later in the semesters. Once assigned together, roommates lived together until their graduation. Moving out or changing roommates was very rare on a few occasions (see Supplementary Information Section 1.2 for more details).

The dataset covers two cohorts of Chinese undergraduates who were admitted by the university in 2011 and 2012, respectively. For each student, we solicited information about their cohort, gender, major, and dorm room, based on which we determined roommate relationships. As a measure of academic performance, we collected the GPA data of these students for the first five successive semesters up to 2014 and further normalized it for each semester to a GPA percentile for students in the same cohort and major (see Supplementary Information Section 1.2 for details). The stable roommate relationship and the longitudinal academic performance data allowed us to study how a student is affected by roommates over time. All students were anonymized in the data collection and analysis process. This study was approved by the Institutional Review Board (IRB) at the University of Electronic Science and Technology of China (IRB No. 1061420210802005).

**Statistical hypothesis test.** Given a tier of classification for students' GPA, following permutation tests[74-76], we perform a statistical test to examine whether the relative ratio $\mathbb{E}$ of each combination (e.g., 1111) in the actual data deviates significantly from its theoretical value 0. Specifically, we generate a roommate null model by implementing the random shuffling process and calculate the null-model relative ratio for each combination: $\widetilde{\mathbb{E}} = (P_n - P_t)/P_t$, where $P_n$ and $P_t$ is the null-model and theoretical frequency of the combination, respectively. By null-model construction, $P_n$ should approach $P_t$, and thus $\widetilde{\mathbb{E}}$ should be close to 0. For each combination, we compare the actual $\mathbb{E}$ with its null-model $\widetilde{\mathbb{E}}$. If $\mathbb{E}$ is significantly above 0, the probability of observing $\mathbb{E} \leq \widetilde{\mathbb{E}}$ in the actual data should be sufficiently small, e.g., less than 0.001. Accordingly, our null hypothesis (H0) is $\mathbb{E} \leq \widetilde{\mathbb{E}}$, and the alternative hypothesis (H1) is $\mathbb{E} > \widetilde{\mathbb{E}}$. To empirically test H0, we generate 1000 roommate null models (where each null model is an independent implementation of the random shuffling process) and calculate $\widetilde{\mathbb{E}}$ under 2-tier, 3-tier, and 4-tier classifications, respectively. We find that $\mathbb{E}$ of some combinations is larger than $\widetilde{\mathbb{E}}$ for all 1000 roommate null models, allowing us to reject H0 and support H1 (i.e., $\mathbb{E}$ is significantly larger than 0 with a *P* value < 0.001 in the one-sided statistical test; the combination is over-represented in the actual data). Similarly, we test whether $\mathbb{E}$ of a combination is significantly below 0. Under the 2-tier classification, for example, combinations with significantly positive $\mathbb{E}$ include 1111 and 2222 (*P* value < 0.001) and those with significantly negative $\mathbb{E}$ include 1112 and 1122 (*P* value < 0.001) as well as 1222 (*P* value < 0.05; see Supplementary Table 2 for the statistical testing results for each combination under these tier classifications). Overall, we find that significantly positive combinations have the same or nearby tier numbers and significantly negative ones have distant tier numbers.

To perform a single statistical test for all combinations together given a tier of classification, we calculate the total relative ratio $\sum|\mathbb{E}|$ and $\sum|\widetilde{\mathbb{E}}|$ by summing up the absolute $\mathbb{E}$ and $\widetilde{\mathbb{E}}$ of each combination, respectively. As $\widetilde{\mathbb{E}}$ is close to 0, $\sum|\widetilde{\mathbb{E}}|$ should also be close to 0. If we assume $\sum|\mathbb{E}| \leq \sum|\widetilde{\mathbb{E}}|$, it is naturally that $\sum|\mathbb{E}|$ is close to 0, yielding $\mathbb{E}$ to be close to 0. There, $\mathbb{E}$ and $\widetilde{\mathbb{E}}$ wouldn't have a significant difference because they are both close to 0. Thereby, to say $\mathbb{E}$ is significantly different from $\widetilde{\mathbb{E}}$, the probability of observing $\sum|\mathbb{E}| \leq \sum|\widetilde{\mathbb{E}}|$ should be sufficiently small, e.g., less than 0.001. Accordingly, our null hypothesis (H0) is $\sum|\mathbb{E}| \leq \sum|\widetilde{\mathbb{E}}|$, and the alternative hypothesis (H1) is $\sum|\mathbb{E}| > \sum|\widetilde{\mathbb{E}}|$. We find that, under 2-tier, 3-tier, and 4-tier classifications, $\sum|\mathbb{E}|$ is always larger than $\sum|\widetilde{\mathbb{E}}|$ for all 1000 roommate null models, allowing us to reject H0 and support H1 with a *P* value < 0.001 (i.e., the overall $\mathbb{E}$ of all combinations is different from 0). Taken together, our hypothesis testing results suggest that $\mathbb{E}$ of some combinations in the actual data deviates significantly from 0, where those with nearby tier numbers are more likely to be observed and those with distant tier numbers are less likely to be observed than random chance, suggesting significant roommate peer effects (see Supplementary Information Section 3.2 for details).

**Regression model.** We employ an ordinary least squares (OLS) model to study the relationship between a student's future performance (GPA_Post) and the average prior performance of their roommate (RM_Avg) and how this relationship is moderated by the differences in roommate prior performance (RM_Diff). The OLS model includes several controls on student demographics and prior performance. Specifically, the empirical specification is given by

$$G_i^{s+1} = b_0 + b_1 G_i^s + b_2 RA_i^s + b_3 RD_i^s + b_4 RA_i^s \times RD_i^s \\ + b_5 D^{Ge} + b_6 D^{Ma} + b_7 D^{Co} + b_8 D^{Se} + \epsilon_i, \qquad (4)$$

where $\epsilon_i$ is the error term for student *i*, and the semester index *s* ranges from 1 to 4. The dependent variable $G_i^{s+1}$ is the student's GPA in semester *s*+1 (GPA_Post), and the independent variable of interest $G_i^s$ is the student's GPA in semester *s* (GPA_Prior). The variable $RA_i^s$ is the roommate average GPA in semester *s* (RM_Avg), $RD_i^s$ is the differences in roommate GPAs in semester *s* (RM_Diff), and $RA_i^s \times RD_i^s$ is their interaction term. The variable $D^{Ge}$ is a gender dummy, which is coded as 1 and 0 for females and males, respectively. The variables $D^{Ma}$, $D^{Co}$, and $D^{Se}$ are major, cohort, and semester dummies, respectively (see Supplementary Table 3 for details).

Moreover, we employ an OLS model to study the relationship between a student's in-dorm ordinal rank (OR_InDorm) according to prior performance and their future performance after controlling their prior performance, the average and differences in roommate prior performance, their gender, major, cohort, and semester. Meanwhile, we examine how this relationship is moderated by other factors including peer heterogeneity. Specifically, the empirical specification is given by

$$G_i^{s+1} = b_0 + b_1 G_i^s + b_2 OR_i^s + b_3 RA_i^s + b_4 RD_i^s + b_5 OR_i^s \times G_i^s + b_6 OR_i^s \times RA_i^s \\ + b_7 OR_i^s \times RD_i^s + b_8 D^{Ge} + b_9 D^{Ma} + b_{10} D^{Co} + b_{11} D^{Se} + \epsilon_i, \qquad (5)$$

where $OR_i^s$ is the OR_InDorm of student *i* in semester *s* (ranging from 1 to 4) and $\epsilon_i$ is the error term. The interaction terms are $OR_i^s \times G_i^s$ between OR_InDorm and GPA_Prior, $OR_i^s \times RA_i^s$ between OR_InDorm and RM_Avg, and $OR_i^s \times RD_i^s$ between OR_InDorm and RM_Diff for student $i$ in semester $s$. All other controls are the same as above (see Supplementary Information Section 5 for details on these variables and Supplementary Table 3 for summary statistics).

**Data Availability**

All data necessary to replicate the statistical analyses and main figures are available in Supplementary Information and have been deposited in the open-access repository Figshare (https://doi.org/10.6084/m9.figshare.25286017)[92]. The raw data of anonymized student accommodation and academic performance are protected by a data use agreement. Those who are interested in the raw data may contact the corresponding authors for access after obtaining Institutional Review Board (IRB) approval.

**Code Availability**

All code necessary to replicate the statistical analyses and main figures has been deposited in the open-access repository Figshare (https://doi.org/10.6084/m9.figshare.25286017)[92].

**Acknowledgements**

The authors thank Min Nie, Defu Lian, Zhihai Rong, Huaxiu Yao, Yifan Wu, Lili Miao, and Linyan Zhang for their valuable discussions. This work was partially supported by the National Natural Science Foundation of China Grant Nos. 42361144718 and 11975071 (T.Z.) and the Ministry of Education of Humanities and Social Science Project Grant No. 21JZD055 (T.Z.).

**Author Contributions Statement**

T.Z. and J.G. designed research; T.Z. collected data; Y.C. and J.G. performed research; Y.C., T.Z., and J.G. analyzed data; J.G. wrote the paper; Y.C. and T.Z. revised the paper.

**Competing Interests Statement**

The authors declare no competing interests.

# Tables

**Table 1. Summary of regression results on the relationship between a student's post-GPA and the average prior GPA of their roommates.**

| Variables | Dependent variable: GPA_Post | | | | |
|---|---|---|---|---|---|
| | (1) | (2) | (3) | (4) | (5) |
| RM_Avg | 0.365*** | 0.050*** | | 0.050*** | 0.042*** |
| | (0.011) | (0.007) | | (0.007) | (0.009) |
| RM_Diff | | | 0.009 | 0.007 | 0.007 |
| | | | (0.009) | (0.009) | (0.009) |
| Interaction: RM_Avg × RM_Diff | | | | | −0.089* |
| | | | | | (0.052) |
| GPA_Prior | | 0.801*** | 0.809*** | 0.801*** | 0.801*** |
| | | (0.004) | (0.004) | (0.004) | (0.004) |
| D (gender) | | 0.039*** | 0.043*** | 0.039*** | 0.039*** |
| | | (0.004) | (0.004) | (0.004) | (0.004) |
| D (semester) | No | Yes | Yes | Yes | Yes |
| Major FE | No | Yes | Yes | Yes | Yes |
| Cohort FE | No | Yes | Yes | Yes | Yes |
| Observations | 15,680 | 15,680 | 15,680 | 15,680 | 15,680 |
| Adjust $R^2$ | 0.059 | 0.668 | 0.667 | 0.668 | 0.668 |
| RMSE | 0.284 | 0.169 | 0.169 | 0.169 | 0.169 |

*Notes:* Independent variables are mean-centered before being included in the regression models except for gender, semester, major, and cohort dummies. Females are in the treatment group. Robust standard errors are reported in parentheses. Significant levels: *$P<0.1$, **$P<0.05$, ***$P<0.01$.

**Table 2. Summary of regression results on the relationship between a student's post-GPA and their in-dorm ordinal rank according to prior GPA.**

| Variables | Dependent variable: GPA_Post | | | | |
|---|---|---|---|---|---|
| | (1) | (2) | (3) | (4) | (5) |
| OR_InDorm | 0.011*** | 0.006** | 0.006** | 0.006** | 0.003 |
| | (0.002) | (0.002) | (0.002) | (0.002) | (0.003) |
| RM_Avg | | 0.030*** | 0.030*** | 0.030*** | 0.035*** |
| | | (0.010) | (0.010) | (0.010) | (0.010) |
| RM_Diff | | 0.007 | 0.007 | 0.008 | 0.007 |
| | | (0.009) | (0.009) | (0.009) | (0.009) |
| Interaction 1: OR_InDorm × GPA_Prior | | | −0.001 | | |
| | | | (0.004) | | |
| Interaction 2: OR_InDorm × RM_Avg | | | | 0.004 | |
| | | | | (0.006) | |
| Interaction 3: OR_InDorm × RM_Diff | | | | | −0.022** |
| | | | | | (0.009) |
| GPA_Prior | 0.840*** | 0.820*** | 0.820*** | 0.820*** | 0.812*** |
| | (0.006) | (0.009) | (0.009) | (0.009) | (0.010) |
| D (gender) | 0.040*** | 0.039*** | 0.039*** | 0.039*** | 0.039*** |
| | (0.004) | (0.004) | (0.004) | (0.004) | (0.004) |
| D (semester) | Yes | Yes | Yes | Yes | Yes |
| Major FE | Yes | Yes | Yes | Yes | Yes |
| Cohort FE | Yes | Yes | Yes | Yes | Yes |
| Observations | 15,680 | 15,680 | 15,680 | 15,680 | 15,680 |
| Adjust $R^2$ | 0.668 | 0.669 | 0.669 | 0.669 | 0.669 |
| RMSE | 0.169 | 0.169 | 0.169 | 0.169 | 0.169 |

*Notes:* Independent variables are mean-centered before being included in the regression models except for gender, semester, major, and cohort dummies. Females are in the treatment group. Robust standard errors are reported in parentheses. Significant levels: *$P<0.1$, **$P<0.05$, ***$P<0.01$.

# Figures

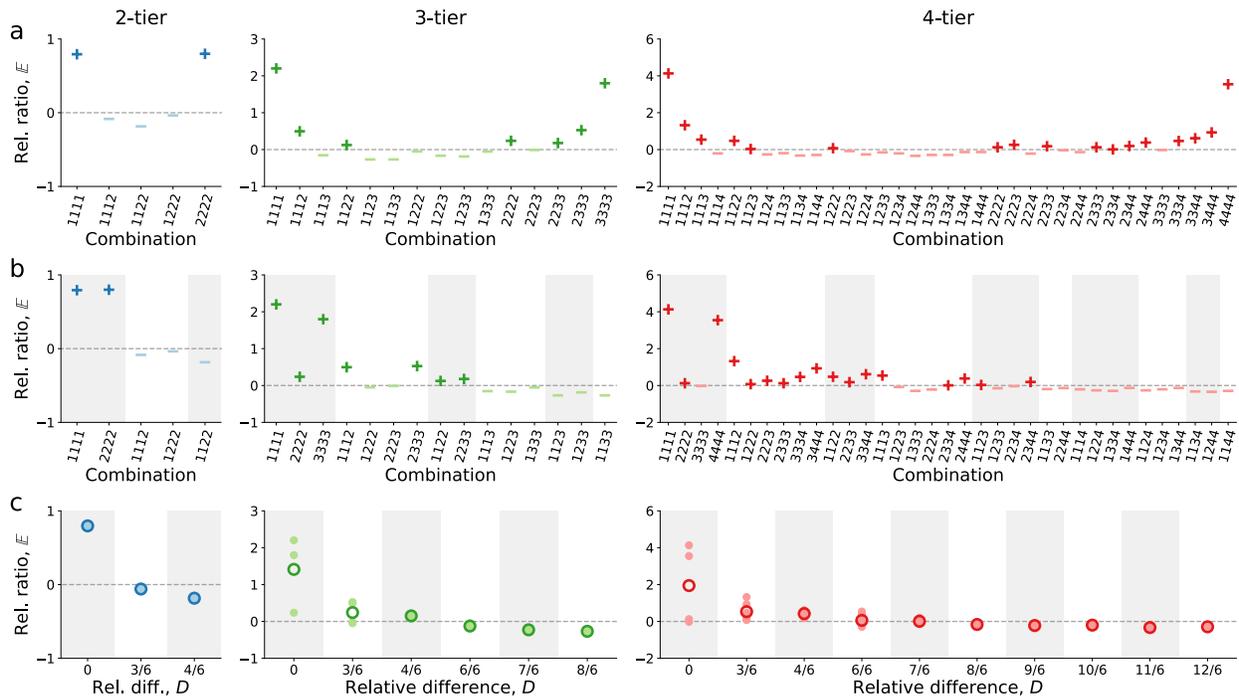

**Fig. 1: The combinations of roommate tiers in a 4-person dorm room.** **(a)** The relative ratio $\mathbb{E}$ of each combination under the 2-tier, 3-tier, and 4-tier classification of GPA, respectively. The x-axis shows all unique combinations in ascending order of tier numbers under a tier classification, and the y-axis shows the relative ratio $\mathbb{E}$ that compares the actual frequency of a combination with its theoretical value. The horizontal dashed line marks 0. Positive and negative $\mathbb{E}$ is marked by '+' and '-', respectively. **(b)** Combinations in ascending order of the relative difference $D$, which measures the average pairwise difference between tier numbers of a combination. The staggered shade marks a group of combinations with the same $D$. **(c)** The negative relationship between the relative ratio $\mathbb{E}$ and the relative difference $D$ based on the actual data. Data points show the $\mathbb{E}$ for each combination, and the hollow circle shows the mean $\mathbb{E}$ for each group with the same $D$.

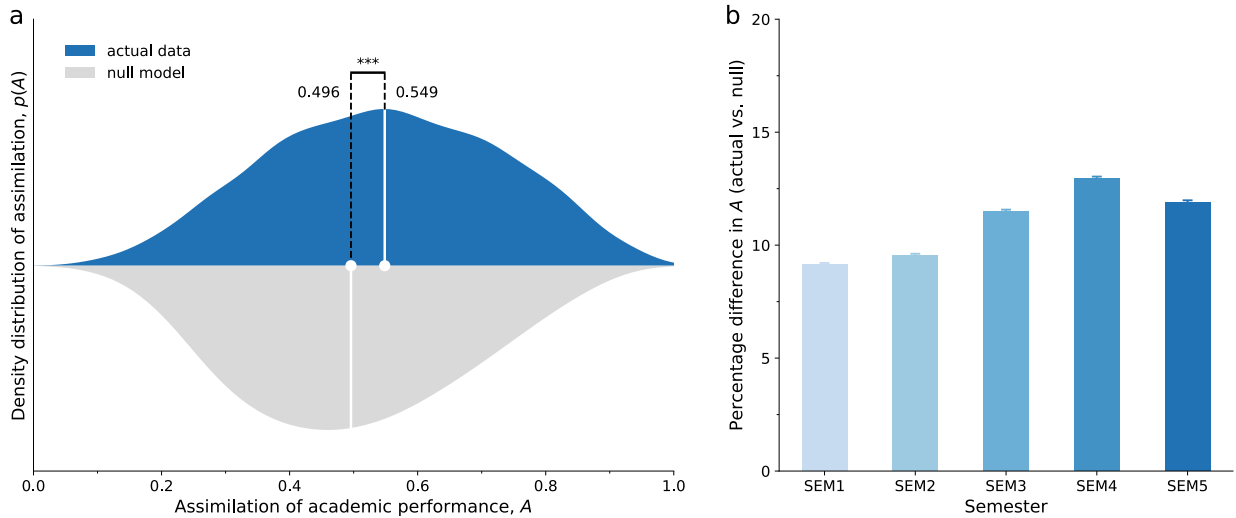

**Fig. 2: The assimilation of roommate academic performance. (a)** The density distribution $p(A)$ of assimilation $A$ for all dorm rooms. Larger assimilation means roommates have more similar academic performance. The left half (in blue) shows the actual assimilation, and the right half (in gray) shows the null-model assimilation. Horizontal dashed lines mark the statistically different means of the two assimilation distributions based on a student's $t$-test ($^{***}P$ value < 0.001). The mean actual assimilation is 10.7% larger than the mean null-model assimilation, which is close to its theoretical value 0.5. The plot is based on the data from all five semesters. **(b)** The overall increasing trend in the actual assimilation from semester 1 to semester 5. The y-axis shows the percentage difference between the mean actual assimilation and the mean null-model assimilation. Error bars represent standard errors clustered for 100 times of independent implementations.

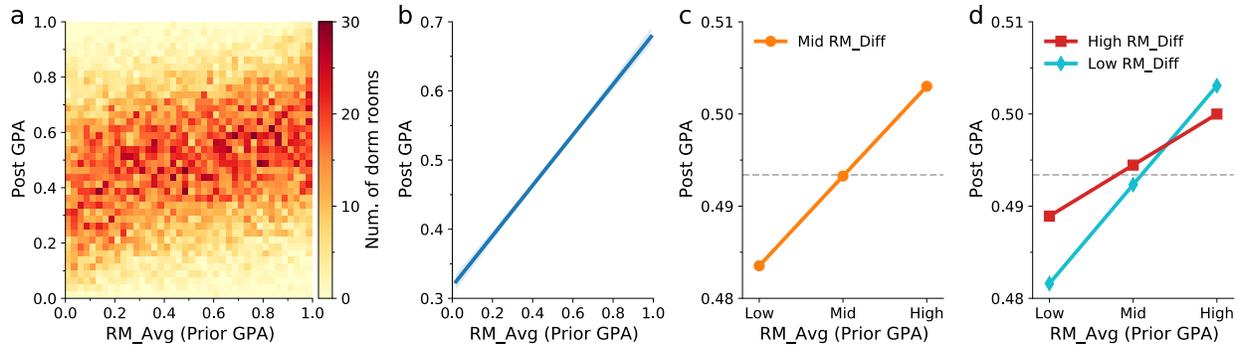

**Fig. 3: The effects of roommate academic performance. (a)** The two-dimensional histogram shows the distributions of dorm rooms on the "GPA_Post – RM_Avg" plane. The y-axis shows the student's post-GPA (GPA_Post), and the x-axis shows the average prior GPA of roommates (RM_Avg). It shows a positive correlation between GPA_Post and RM_Avg (Pearson's $r = 0.244$; $P$ value < 0.001). **(b)** The regression plot for the relationship between GPA_Post and RM_Avg (center line) with the 95% confidence intervals (error bands), where the model includes no controls. **(c)** The plot for the relationship between GPA_Post and RM_Avg, where the model includes controls and fixed effects (see Table 1 for details). The "Low" and "High" on the x-axis represent 1 standard deviation (*s.d.*) below and above the mean ("Mid") of RM_Avg, respectively. The horizontal dashed line marks the regression constant. **(d)** The plot for the moderating effects. The relationship between GPA_Post and RM_Avg is moderated by the differences in roommate prior GPAs (RM_Diff). The "Low" and "High" in the legend represent 1 *s.d.* below and above the mean ("Mid") of RM_Diff, respectively. The horizontal dashed line marks the regression constant.

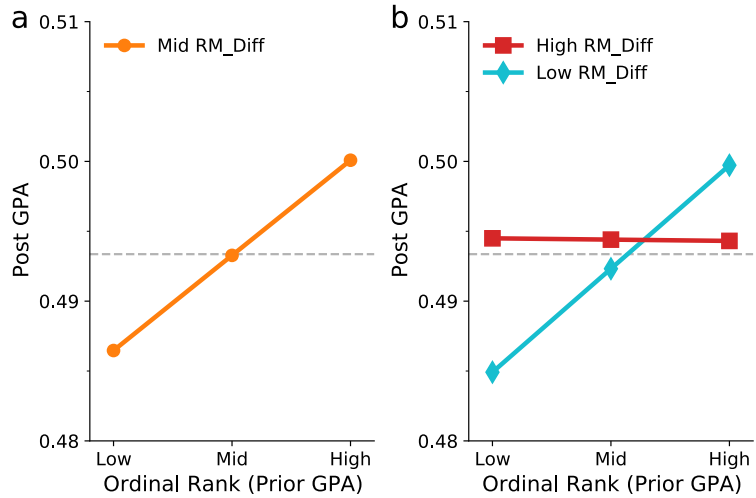

**Fig. 4: The effects of in-dorm ordinal rank. (a)** The plot for the relationship between a student's GPA in the current semester (GPA_Post) and their ordinal rank according to GPA in the previous semester (OR_InDorm), where a larger rank value corresponds to a lower GPA. The OLS regression model includes controls and fixed effects (see Table 2 for details). The "Low" and "High" on the x-axis represent 1 standard deviation (*s.d.*) below and above the mean ("Mid") of OR_InDorm, respectively. The horizontal dashed line marks the regression constant. **(b)** The plot for the moderating effects. The relationship between GPA_Post and OR_InDorm is moderated by the differences in roommate GPAs in the previous semester (RM_Diff). The "Low" and "High" in the legend represent 1 *s.d.* below and above the mean ("Mid") of RM_Diff, respectively. The horizontal dashed line marks the regression constant.

# Supplementary Information:

# Heterogeneous peer effects of college roommates on academic performance

Yi Cao[1,2], Tao Zhou[1,2,*], Jian Gao[3,4,5,6,*]

[1] CompleX Lab, University of Electronic Science and Technology of China, Chengdu, China
[2] Big Data Research Center, University of Electronic Science and Technology of China, Chengdu, China
[3] Center for Science of Science and Innovation, Northwestern University, Evanston, IL, USA
[4] Kellogg School of Management, Northwestern University, Evanston, IL, USA
[5] Northwestern Institute on Complex Systems, Northwestern University, Evanston, IL, USA
[6] Faculty of Social Sciences, The University of Hong Kong, Hong Kong SAR, China
[*] Correspondence to zhutou@ustc.edu or jian.gao1@kellogg.northwestern.edu

# Table of Contents

# 1. Data on student roommate and academic performance

We collect the accommodation and academic performance data of undergraduates from a public research-intensive university in China. In the university, students live in identical 4-person dorm rooms during their entire undergraduate program after a one-time plausible random assignment of roommates upon registration before the first semester. This randomization allows us to study the impact of college roommate peers on academic performance. In the following, we first introduce the differences in the layouts of a typical student dorm room in Western universities and in Chinese universities. We then introduce the roommate assignment process in the Chinese university and the longitudinal datasets of student accommodation and academic performance.

## 1.1 Dorm rooms in Western and Chinese universities

An on-campus dormitory, or residence hall, usually contains a large number of dorm rooms, which provide a place for students to sleep, offering opportunities for their educational and personal development[1-7]. Different from open and dynamic education environments such as classrooms[8-11], dorm rooms provide a close-knit living environment for students to interact and potentially learn from each other[12-14]. Arguably, while dorm rooms are not the primary places in which learning takes place such as classrooms and university libraries, they offer a highly interpersonal and spillover environment for a small group of student peers. This non-learning-oriented dormitory environment is ideally suited for studying the impact of student peers on learning outcomes.

In most Western countries, first-year undergraduates are required to live in on-campus dormitories for a certain period of time[14-18]. After this period, students have the option to continue living on-campus for the next academic year or to live off-campus. In the Western university dormitory, a typical dorm suite contains several bedrooms, where each student occupies a single bedroom (where a room contains only one bed) and shares public areas with other suite mates including living room, kitchen, and bathroom. In Supplementary Figure 1a, we show the floor plan of a typical Western university dorm suite containing four bedrooms and public areas. It is uncommon for a bedroom within a dorm suite to contain two or more beds and for students to have roommates.

Different from many Western universities, almost all universities in China provide on-campus dormitories for undergraduate students for their entire period of study, which is four years for most undergraduate programs[19-22]. Only a few students with special permission live off-campus. There are also substantial differences in the layouts of dorm rooms between the two. In the Chinese university dormitory, a typical dorm room contains a single bedroom with several beds, where each student occupies one bed and shares public areas in the dorm room with other roommates. In Supplementary Figure 1b, we show the floor plan of a typical Chinese university dorm room containing four beds and public areas including a balcony and washroom. The Chinese dorm room is much smaller than the dorm suite in Western universities, and it usually contains a larger number of beds, usually ranging from 2 beds to as many as over 20 beds in one single room, whereas 4 beds in a room for undergraduates are the most common setup in Chinese universities. This tight layout is in part due to limited resources but huge demands for on-campus accommodations. For example, at the university under study, there are about 5,000 students in each cohort (e.g., those who were admitted by the university in the same year) in the undergraduate programs, meaning that the university has to offer in total of about 20,000 beds for four cohorts of undergraduates. Dorm rooms in the same dormitory building are identical to each other. New students are randomly assigned to dorm rooms, and they live together for four years until graduation (see Section 1.2 for details). Taken together, a Chinese dorm room provides a more interactive environment than a Western dorm suite, and it helps generate more interpersonal interactions between roommates.

Moreover, different from Western universities, where on-campus student residential buildings are integrated into local communities, most universities in China have clear geographical boundaries protected by fences, and university gates are guarded. In the university under study, dorm buildings are all within the university boundary (i.e., on campus), and they are identical and geographically very close (see Supplementary Figure 2 for a sketch of the campus layout). This suggests that dorm rooms should have the same or very similar environment.

**1.2 Roommate assignment in university dorms**

In Western countries, most universities provide on-campus dormitories to undergraduate students for the first year or the first six quarters of enrollment[14-18]. To live on campus, new students have to fill out housing applications to the university housing management office and sign a housing contract with the office. Throughout the application, students have the opportunity to choose a dormitory suite and provide their lifestyles and living preferences for consideration. The housing assignment team reviews each student's personal preferences and uses them for dorm and suite mate matching[18,23]. If students provide no such personal preferences, their dorm and suite mates will be randomly assigned[12,24-27]. Although not all personal preferences are guaranteed to be matched[28], the suitemate assignment is not completely random because students with similar lifestyles and living preferences have a higher probability of living together.

By contrast, in almost all Chinese universities, the assignment of roommates and dorm rooms is plausibly random[13,22,29-31]. Taking the university under study as an example, we consulted the university administrators from the Housing Management Office and the Student Admission Office, yielding three pieces of evidence supporting the random assignment.

First, the initial assignment followed university student accommodation management regulations, and there was no consideration of student personal preferences. Before student registration in the first semester, the housing office received a list of incoming students with the information of their name, gender, and student ID. Detailed departmental information is coded in the student ID (e.g., "2008060205003," a hypothetical student ID number to illustrate its structure; the same below), where the first four digits represent cohort (e.g., "2008", namely, the year of admission), each of the following two digits represents the school (e.g., "06": School of Computer Science and Engineering), major (e.g., "02": Information Security major in the school), and administrative unit (e.g., "05": the 5th unit in the major; one unit contains about 30 students), and the last three digits represent the student's order in their administrative unit (e.g., "003": the 3rd student in the unit). Upon receiving the student list, the housing office separated female and male students because there was no gender mix in dorm rooms or buildings and then assigned them to dorm rooms in order of student ID. There, students of the same gender, administrative unit, major, or school were assigned to the same 4-person dorm room, on the same floor, and in the same dorm building when possible. For example, students with student ID numbers "2008060205001," "2008060205002," "2008060205003," and "2008060205004" were assigned together. When an administrative unit had remaining students, they were paired with students of the same gender in the next unit, the same major, and the same school when possible. Our analysis of the accommodation data showed that the university's student accommodation management regulations were enforced on the initial roommate assignment of roommates. Specifically, we found that roommates tend to have nearby student IDs (Supplementary Figure 3a), a pattern that is more pronounced in the actual data than in the roommate null model (see Section 3.1 for details), and roommates in most dorm rooms have the same major (Supplementary Figure 3b). Moreover, it was confirmed that the housing office neither had access to students' prior academic performance or other personal data nor allowed them to choose roommates or dorm rooms based on personal preferences. Students were informed of their accommodation information only when they checked in before the first semester.

Second, the initial assignment did not depend on the student's prior academic performance. We consulted the admission office to understand the relationship between student IDs and their academic performance before college enrollment. We were informed that there were no simple, one-size-fits-all criteria to compare the ability of students who were admitted into the same major. The assignment of student IDs was plausibly random for the following reasons. (1) The university used a quota system to admit students into different majors, and students were admitted in the order of majors to which they applied. The quotas for different provinces or regions were allocated according to the university's discretion and preferences, whereas the university-hosting province received more quotas. (2) The National Higher Education Entrance Examination (i.e., "Gaokao" in China) was held by provincial governments using different evaluation systems. The total scores of the exams were hardly comparable across different systems (not like the total SAT score in the US, for example). Therefore, no simple criteria were obvious enough for administrators to evaluate students' academic performance before college enrollment. (3) The admission office randomly assigned students in the same major to one or more administrative units (about 30 students in each administrative unit) and assigned a student ID number to each student (see above for details) without considering their prior academic performance. There was no mix of majors in the same administrative unit. For students in the same major, the only obvious consideration was gender. Specifically, in each administrative unit, male students were always assigned smaller student ID numbers (e.g., "2008060205003") than female students (e.g., "2008060205029").

Third, there was no significant selection process later in the semesters. Once assigned to a dorm room by the housing office, four students lived together until their graduation. There was no systematic reassignment of roommates or change of dorm rooms after the initial assignment. Moving out of a dorm room or from one dorm room to the other should be very rare on a few occasions. All applications for a dorm room change or roommate change were reviewed by the housing office. Exceptions were only given in justified circumstances, such as when a student transferred to a different program, applied for sick leave or had irreconcilable conflicts with other roommates. According to the housing office, these circumstances were very rare at the time.

Taken together, the initial assignment is plausibly random concerning students' prior academic performance and preferences. While we lacked data on student prior performance and personal data to directly verify the random assignment, our consultation with university administrators gave us extra confidence in assuming that the initial assignment of roommates and dorm rooms was plausibly random because the administrative processes involved no obvious consideration of student prior academic performance, socioeconomic background, or personal preference.

**1.3 Student accommodation and academic performance data**

From the university under study, we collect longitudinal data on both the accommodation and the academic performance of Chinese undergraduate students. The data covers the 2011 and 2012 cohorts of undergraduates, who were randomly assigned to identical 4-person dorm rooms before the first semester and lived together until their graduation. There are no foreign students in the data. For accommodation, we collect each student's dorm room number. Students with the same dorm room number are identified as roommates, and this roommate relationship is static by construction. For academic performance, we collect the grade point average (GPA) for each student and semester in 2011-2014. The GPA data covers the first five consecutive semesters for the 2011 cohort and the first three consecutive semesters for the 2012 cohort. We exclude those students whose GPA information was not available at the time of data collection. For comparisons across different majors and cohorts, we transform each student's GPA into a percentile $R$ among students in the same major and cohort for each semester, where $R = 0$ and $R = 1$ correspond to the lowest and highest academic performance, respectively. We inferred the major of each student from their student ID (see Section 1.2 for details). We dropped those for which we couldn't draw a confident inference or that had missing values in other variables. There are in total 36 majors for the two cohorts of students (34 majors for one cohort and 31 majors for the other). Most majors have 55~150 students, and majors with a few students are rare (Supplementary Figure 4). At the minimum, one major has fifteen students. Together, the data collection and cleaning process led to a sample of 5,272 students, who lived in identical 4-person dorm rooms for up to five semesters.

## 2. Tier combinations of roommate academic performance

To study roommate peer effects, we start by analyzing the composition of students in a typical 4-person dorm room in terms of their academic performance. Specifically, we divide students into equal-sized tiers based on their GPA percentiles $R$ (hereafter GPA for simplicity) and compare the actual probability of observing a tier combination of roommates in the real data with its theoretical probability. In the following, we first introduce the theoretical probability of each tier combination under different tiers of classification (e.g., 3 tiers and 4 tiers). We then introduce the comparison between the actual and theoretical probabilities, followed by the empirical results for each semester.

### 2.1 The theoretical probability for each combination

Given a tier of classification, students are categorized into equal-sized tiers based on their GPA, where a larger tier number indicates better academic performance. For example, under the 2-tier classification, a student with $R = 0.7$ (i.e., GPA is higher than 70% of students in the same major) is in Tier 1. The tier number of the same student may change under different classifications. For example, the student is in Tier 3 under the 4-tier classification. Under a given tier of classification, each 4-person dorm room has a tier combination without particular order, for example, 3444 (i.e., one student is in Tier 3, and the other three students are in Tier 4) is identical to 4344, 4434, and 4443. Therefore, we use the one in ascending tier order to delegate all identical others. Under the 2-tier classification, there are five combinations (1111, 1112, 1122, 1222, and 2222). The numbers are 15 and 35 under the 3-tier and 4-tier classifications, respectively (see Supplementary Figure 5 for a list of all combinations under these classifications). Throughout the study, we only consider the setting of a typical 4-person dorm room (see Section 1.1 for details), for simplicity and consistency with the case of the university under study.

If there is no roommate peer effect, each roommate's GPA should be independent and identically distributed (i.i.d.), and the tier numbers of roommates determined by their GPA and the tier classification used should also be independent of each other. Therefore, we can calculate the theoretical probability $P_t$ for each combination. It should be noted that the theoretical probabilities of different combinations are not always the same, and they shouldn't be directly compared with each other because the total numbers of identical combinations delegated by the combination in ascending order are sometimes different. For a simple example, under the 2-tier classification, the theoretical probability $P_t$ of combination 1112 is calculated by

$$P_t(1112) = C_4^1 \left(\frac{1}{2}\right)^3 \left(\frac{1}{2}\right) = \frac{1}{4}. \qquad (1)$$

The value is four times as large as theoretical probability $P_t$ of combination 1111 calculated by

$$P_t(1111) = \left(\frac{1}{2}\right)^4 = \frac{1}{16}. \qquad (2)$$

The differences in the theoretical probabilities are because of combination 1112 delegates four identical combinations (namely, 1112, 1121, 1211, and 2111), while combination 1111 only delegates one identical combination. In Supplementary Table 1, we show all combinations and their theoretical probabilities under the 2-tier, 3-tier, and 4-tier classifications, respectively.

## 2.2 Comparison between actual and theoretical probabilities

The actual probability $P_a$ of observing a combination in the actual data is calculated by the share of dorm rooms with the combination. If there are significant roommate peer effects on academic performance, the GPAs of roommates should be correlated, leading to a higher or lower probability of observing some combinations across all dorm rooms. As mentioned above, however, the actual probabilities of different combinations shouldn't be directly compared with each other because their theoretical probabilities are not always the same. This results in difficulties in assessing whether a combination is over-represented or under-represented in the real data only by looking at the face values of actual probabilities. To address this challenge, we introduce a metric named relative ratio $\mathbb{E}$, which is calculated by the relative difference between the actual probability and the theoretical probability of the same combination. Formally, the relative ratio $\mathbb{E}$ is given by

$$\mathbb{E} = \frac{P_a - P_t}{P_t}, \qquad (3)$$

where $P_a$ and $P_t$ are the actual probability and the theoretical probability of the same combination, respectively. A positive (negative) relative ratio $\mathbb{E}$ means the combination is more (less) likely to be observed in the actual data than expected by chance. In the main text (Fig. 1), we have shown the relative ratio $\mathbb{E}$ of combinations under 2-tier, 3-tier, and 4-tier classifications for a 4-person dorm room based on the data for all five semesters together. Here we further show that the main conclusions are largely robust when using the data for each semester. Specifically, Supplementary Figure 5 presents all possible combinations in ascending order of tier numbers under 2-tier, 3-tier, and 4-tier classifications and the relative ratio $\mathbb{E}$ of each combination from semester 1 to semester 5, where each column corresponds to the data for one semester.

Analyzing the data, we notice that some combinations tend to have a larger relative ratio $\mathbb{E}$ than others, and this pattern appears to hold for all five semesters (Supplementary Figure 5). The result suggests that the actual probabilities of these combinations are larger than their theoretical probabilities. Moreover, we find that these combinations tend to contain nearby tier numbers such as 1111, and 1112, suggesting that roommates tend to have similar academic performance. This observation prompts us to further explore the relationship between the tier heterogeneity of a combination (i.e., the differences in the tier numbers) and its relative ratio $\mathbb{E}$. More specifically, we first calculate the relative difference $D$ in the tier numbers of each combination:

$$D = \frac{1}{6}\sum_{u \neq v}|l_u - l_v|, 1 \leq u < v \leq 4, \qquad (4)$$

where $l_u$ and $l_v$ is the tier number of roommates $u$ and $v$, respectively. A smaller relative difference $D$ means that roommates have closer tier numbers and thus more similar academic performance. In Supplementary Figure 6, we group combinations with the same *D* and rearrange the order of combinations in the descending order of *D* and tier numbers. We find that there appears to be a negative association between the relative ratio $\mathbb{E}$ and the relative difference *D*, where $\mathbb{E}$ is larger for combinations with smaller *D*. This pattern is more pronounced when we take an average of $\mathbb{E}$ for all combinations with the same *D*. Specifically, Supplementary Figure 7 presents a much clear positive association between $\mathbb{E}$ and *D* for each of the five semesters. This result suggests a higher probability of observing a dorm room in which roommates have smaller differences in their academic performance in the actual data than its theoretical value. Together, these results show that roommates are more similar in academic performance, suggesting the presence of roommate peer effects (see Section 3.2 for detailed statistical testing results).

## 3. Roommate null model and statistical test

We perform a statistical hypothesis test to further demonstrate the presence of roommate peer effects on academic performance by comparing observations from the actual data with those from a roommate null model. In the following, we first introduce the method for constructing the roommate null model based on the actual accommodation data. We then introduce the method for the statistical hypothesis testing and the results for quantifying roommate peer effects.

### 3.1 Construction of a roommate null model

The null model is a widely used approach to studying non-trivial features of complex systems[32-34]. An appropriate null model of a complex system not only satisfies a collection of constraints, for instance, a network null model that preserves degree distributions of the original network[35-37], but also provides an unbiased random structure of other features, offering a baseline to examine whether displayed features of interest in the original complex system are truly non-trivial. Over the past few decades, null-model approaches have been applied to test causal effects in many complex social systems[38-40]. For instance, in the social network literature, randomizations as one type of null models are used to study the impact of network interventions on social relationships[41]. Building on traditional statistical methods and inspired by recent applications of null-model approaches, we construct a roommate null model to test whether roommates are more similar in their academic performance than expected by chance alone.

Ideally, we would want the roommate null model to replicate the actual random assignment of roommates, and at the same time, it takes into account any non-random aspects of the actual assignment (see Section 1.2 for details). To build a roommate null model, we start with the actual roommate configuration in the real data and randomly shuffle students between dorm rooms while preserving their compositions of cohort, gender, and major. Specifically, in practice, each time we randomly select two students in the same cohort, gender, and major and then swap them into each other's dorm room. Besides these controlled characteristics during the random shuffling process, the two students should have random roommates concerning their academic performance and other characteristics. We repeat the student random selection and swapping many times, for example, over 100,000 times to produce one implementation. By repeating this entire process, we construct a plausible roommate null model that consists of 1000 independent implementations. We show that the roommate null model that we construct successfully preserves these controlled student characteristics. For example, the major compositions of all dorm rooms in the roommate null model are the same as in the actual data (Supplementary Figure 3b).

## 3.2 Statistical hypothesis testing

A statistical test is a powerful tool for examining whether the empirical data sufficiently supports a particular hypothesis that is tested. In the main text (Fig. 1), we find that the relative ratio $\mathbb{E}$ of different combinations varies and deviates from 0. For example, $\mathbb{E}$ of combination 1111 is positive and much larger than others such as combination 1122. Here we further perform a statistical test to examine whether a combination's actual relative ratio $\mathbb{E}$ deviates significantly from its theoretical value which is 0, in other words, whether the actual probability $P_a$ is significantly different from its theoretical probability $P_t$. Considering that $P_t$ of a combination is only one value, making it hard to tell whether $P_a$ is significantly larger or smaller, we employ the roommate null model to perform a statistical test, where the null-model probability $P_n$ is used to proxy $P_t$.

Specifically, given a tier of classification, we first generate a roommate null model (see Section 3.1 for details) and calculate the null-model relative ratio for each combination:

$$\widetilde{\mathbb{E}} = \frac{P_n - P_t}{P_t}, \tag{5}$$

where $P_n$ and $P_t$ are the combination's null-model and theoretical probabilities, respectively. By the null-model construction, $P_n$ should approach $P_t$, and thus $\widetilde{\mathbb{E}}$ should be close to 0. For each combination, we compare the actual $E$ with its null-model $\widetilde{\mathbb{E}}$. If $\mathbb{E}$ is significantly above 0, the probability of observing $\mathbb{E} \leq \widetilde{\mathbb{E}}$ in the actual data should be sufficiently small, e.g., less than 0.001. Accordingly, our null hypothesis (H0) is $\mathbb{E} \leq \widetilde{\mathbb{E}}$, and the alternative hypothesis (H1) is $\mathbb{E} > \widetilde{\mathbb{E}}$. To empirically test H0, we generate 1000 roommate null models (where each null model is an independent implementation of the random shuffling process) and calculate $\widetilde{\mathbb{E}}$ under 2-tier, 3-tier, and 4-tier classifications, respectively. We find that $\mathbb{E}$ of some combinations is larger than $\widetilde{\mathbb{E}}$ for all roommate null models, allowing us to reject H0 and support H1 (i.e., $\mathbb{E}$ is significantly larger than 0 with a $P$ value < 0.001 in the one-sided statistical test; the combination is over-represented in the actual data). Similarly, we test whether $\mathbb{E}$ of a combination is significantly below 0. Under the 2-tier classification, for example, combinations with significantly positive $\mathbb{E}$ include 1111 and 2222 ($P$ value < 0.001) and those with significantly negative $\mathbb{E}$ include 1112 and 1122 ($P$ value < 0.001) as well as 1222 ($P$ value < 0.05). In Supplementary Table 2, we show the statistical testing results for each combination under 2-tier, 3-tier, and 4-tier classifications, respectively.

To perform a single statistical test for all combinations together given a tier of classification, we calculate the total relative ratio $\sum|\mathbb{E}|$ and $\sum|\widetilde{\mathbb{E}}|$ by summing up the absolute $\mathbb{E}$ and $\widetilde{\mathbb{E}}$ of each combination, respectively. As $\widetilde{\mathbb{E}}$ is close to 0, $\sum|\widetilde{\mathbb{E}}|$ should also be close to 0. If we assume $\sum|\mathbb{E}| \leq \sum|\widetilde{\mathbb{E}}|$, it is naturally that $\sum|\mathbb{E}|$ is close to 0, yielding $\mathbb{E}$ to be close to 0. There, $\mathbb{E}$ and $\widetilde{\mathbb{E}}$ wouldn't have a significant difference because they are both close to 0. Thereby, to say $\mathbb{E}$ is significantly different from $\widetilde{\mathbb{E}}$, the probability of observing $\sum|\mathbb{E}| \leq \sum|\widetilde{\mathbb{E}}|$ should be sufficiently small, e.g., less than 0.001. Accordingly, our null hypothesis (H0) is $\sum|\mathbb{E}| \leq \sum|\widetilde{\mathbb{E}}|$, and the alternative hypothesis (H1) is $\sum|\mathbb{E}| > \sum|\widetilde{\mathbb{E}}|$. We find that, under 2-tier, 3-tier, and 4-tier classifications, $\sum|\mathbb{E}|$ is always larger than $\sum|\widetilde{\mathbb{E}}|$ for 1000 roommate null models, allowing us to reject H0 and support H1 with a $P$ value < 0.001 (i.e., the overall $\mathbb{E}$ of all combinations is different from 0). Taken together, our hypothesis testing results show that $\mathbb{E}$ of some tier combinations in the actual data deviates significantly from 0, where those with nearby tier numbers of academic performance are more likely and those with distant tier numbers are less likely to be observed than random chance, suggesting that roommate peer effects on academic performance are significant.

## 4. Assimilation of roommate academic performance

The positive association between the relative ratio $\mathbb{E}$ and the relative difference $D$ of student tier combinations in dorm rooms suggests that roommates with similar academic performance are more likely to be observed in the real data than expected by chance. To examine this relationship at a more granular level, we develop an assimilation metric to measure the differences in roommate academic performance. In the following, we first introduce the assimilation metric and then introduce the statistical test to demonstrate the presence of roommate peer effects.

### 4.1 Academic performance assimilation

We generalize the analysis of tier combinations to the most granular tier for classification by directly dealing with the original GPA percentile $R \in [0, 1]$ (hereafter, GPA for short), where a larger value means better academic performance. More specifically, we develop an assimilation metric $A$ to quantify the differences in the GPAs between roommates (Supplementary Figure 8). Formally, the assimilation $A$ for a 4-person dorm room is calculated by

$$A = 1 - \frac{1}{4}\sum_{u \neq v}|R_u - R_v|, 1 \leq u < v \leq 4, \tag{6}$$

where $R_u$ and $R_v$ are the GPAs of roommates $u$ and $v$, respectively. The assimilation $A$ is between 0 and 1, where a larger value means that roommates have more similar academic performance. Based on the data, we can calculate the distribution of actual assimilation $A$ of all dorm rooms. If there is no roommate peer effect, the $R$ of each roommate should be independent and identically distributed (i.i.d.), and the value of $R$ follows a uniform distribution in $[0, 1]$. Therefore, we can calculate the expectation (mean value) of the theoretical assimilation $A_t$ for a 4-person dorm room:

$$\langle A_t \rangle = 1 - \frac{3}{2} \times \iint_0^1 |x - y|\, dx\, dy = \frac{1}{2}. \tag{7}$$

To perform a statistical test on whether the actual assimilation is significantly larger than its theoretical value, we use the null-model assimilation that is calculated based on the roommate null model (see Section 3.1 for details) to proxy the theoretical assimilation.

We compare the distribution of the actual assimilation for all dorm rooms with the one under the roommate null model that consists of 1000 independent implementations. For all five semesters together (see Fig. 2a in the main text), we find that the mean actual assimilation in the actual data is 0.549, which is much larger than the mean null-model assimilation in the null model (0.496). When looking at each semester (Supplementary Figure 8a), we find that the main result holds robustly and the mean actual assimilation exhibits an overall increasing trend. Specifically, the mean actual assimilation has steadily increased in the first three semesters (Supplementary Figure 8b), which is consistent with the literature showing that peer effects are strong and persistent when student friendships or peer relationships last over a year[21,42,43], but starts to decrease in the fifth semester, which may be because that senior undergraduates can take different elective courses and have more outside activities that decrease their interactions with roommates[44].

### 4.2 Statistical tests for differences in assimilation

We perform a statistical test on whether the actual assimilation is significantly different from the null-model assimilation. Specifically, we use the student's *t*-test, which is a common application to test whether the means of two distributions are significantly different. Our null hypothesis (H0) is that the two assimilation distributions are from the same distribution. The student's *t*-test performed on the actual data rejects the null hypothesis (H0) as the *P* value is less than 0.001. The result supports that the null-model assimilation and the actual assimilation are from different distributions, and they have different means (student's *t*-test; *P* value < 0.001). Specifically, the mean actual assimilation (0.549) of all dorm rooms is about 10.7% larger than the mean null-model assimilation (0.496; Supplementary Figure 8b), which is very close to its theoretical value of 0.5 in complete randomization. Taken together, these results suggest that roommates exhibit larger assimilation than random chance, demonstrating the presence of roommate peer effects.

### 4.3 Robustness check for the assimilation results

In the main analysis, we compare the means of the actual assimilation distribution with the null-model assimilation distribution, where the student's *t*-test suggests that the two distributions are different. Here we check the robustness of these results using an alternative approach, where we examine the significance of each dorm room's assimilation. Specifically, we first calculate the average assimilation (mean) of all dorm rooms under the roommate null model consisting of 100 independent implementations (see Section 3.1 for details) and its standard error (*s.e.*). We then calculate the assimilation of each dorm room and identify those that have significantly larger assimilation than the average null-model assimilation (*P* value = 0.05; a value that is larger than mean+1.96×*s.e.*, i.e., the upper 95% confidence interval). We find that about 60% of dorm rooms have larger-than-null-model assimilation (Supplementary Figure 9a), and this fraction exhibits an increasing trend as well (Supplementary Figure 9b), suggesting that peer effects overall increase as students live together longer. Our conclusions are supported by two different methods.

### 4.4 The assimilation results by different gender

In the university under study, female and male students were assigned to dorm rooms in different buildings, and there was no gender mix in dorm rooms or buildings. This motivates us to unpack the assimilation analysis by gender and add gender as a control to the roommate null model and the OLS regression model. We find that while female students represent only a small proportion of the student sample (about 23%), they have significantly better academic performance on average than their male counterparts (*P* value < 0.001; Supplementary Figure 10a), which is consistent with the positive coefficient of the gender dummy (where females are in the treatment group) in Table 1 and Table 2 of the main text. Yet, here we find that there is no significant gender difference in the average assimilation either for all five semesters together (*P* value = 0.18; Supplementary Figure 10b) or for each semester (*P* value > 0.1; Supplementary Figure 10c), suggesting that roommate peer effects are prevalent across genders.

## 5. Regression models and regression results

We perform a Granger causality type of analysis to better understand roommate peer effects. Briefly, we use an ordinary least squares (OLS) model to regress a student's GPA in the current semester (GPA_Post) against several different factors including the student's GPA in the previous semester (GPA_Prior), the roommate prior average GPA (RM_Avg), the differences in roommate prior GPAs (RM_Diff), the student's in-dorm ordinal rank according to prior GPA (OR_InDorm), and the dummy variables of the student cohort, gender, major, and semester. Supplementary Table 3 shows the summary statistics of all variables and their symbols. All independent variables are mean-centered before being included in the OLS regression except for dummies.

### 5.1 Regression on the effects of peer heterogeneity

In our first analysis, we employ an OLS model to regress the relationship between a student's future GPA (GPA_Post) and the roommates' average prior GPA (RM_Avg). As we are interested in the effects of peer heterogeneity, we include the differences in the roommates' prior GPAs (RM_Diff) and their interaction with RM_Avg in the OLS model. Other factors may independently affect a student's future GPA. Therefore, we include controls on the student's prior GPA, cohort, gender, major, and semester. Specifically, the empirical specification is given by

$$G_i^{s+1} = b_0 + b_1 G_i^s + b_2 RA_i^s + b_3 RD_i^s + b_4 RA_i^s \times RD_i^s \\ + b_5 D^{Ge} + b_6 D^{Ma} + b_7 D^{Co} + b_8 D^{Se} + \epsilon_i, \tag{8}$$

where $\epsilon_i$ is the error term for student $i$, and the semester index $s$ ranges from 1 to 4. The dependent variable $G_i^{s+1}$ is the student's GPA in semester $s$+1 (GPA_Post), and the independent variable of interest $G_i^s$ is the student's GPA in semester $s$ (GPA_Prior), respectively. The variable $RA_i^s$ is the average GPA of roommates in semester $s$ (RM_Avg), $RD_i^s$ is the differences in roommate prior GPAs (RM_Diff), and $RA_i^s \times RD_i^s$ is their interaction term. The variable $D^{Ge}$ is gender dummy, which is coded as 1 and 0 for female and male students, respectively. The variables $D^{Ma}$, $D^{Co}$, and $D^{Se}$ are major, cohort, and semester dummies, respectively.

The regression results for all five semesters together are summarized in Table 1 of the main text. We find that (1) the average prior GPA of roommates has a significantly positive effect on a student's post-GPA, (2) a student's prior GPA has the strongest effect on a student's post-GPA, and (3) the differences in roommate prior GPAs moderates the relationship between post-GPA and roommate average prior GPA such that the positive effects of roommate average prior GPA on a student's post-GPA are more pronounced when the differences in roommate prior GPAs are small. Fig. 3d of the main text shows the margin plot of the linear relationship between GPA_Post and RM_Avg, which is moderated by RM_Diff. To better assess the significance of the moderating effect, we perform a simple slope testing and summarize the results in Supplementary Figure 11a. We find that the slope decreases slightly with the increase of RM_Diff but it remains positive and statistically different from 0 (*P* value < 0.1). More specifically, the slope *b* = 0.055 (*s.e.* = 0.008; 95% CI = [0.040, 0.070]) when the differences in roommate prior GPAs are low (i.e., 1 *s.d.* below its mean), and the slope *b* = 0.028 (*s.e.* = 0.015; 95% CI = [−0.001, 0.057]) when the differences in roommate prior GPA are high (i.e., 1 *s.d.* above its mean). These results show that the positive relationship between a student's post-GPA and their roommate average prior GPA is more (less) pronounced when the differences in their roommate prior GPAs are low (high), suggesting the significant moderating role of peer heterogeneity on roommate peer effects.

While the regression analysis suggests that the roommate peer effects are significant, the effect size appears to be modest. Specifically, a 100-point increase in the roommate average prior GPA is associated with a 5-point increase in the student's GPA (see Table 1 in the main text). The effect is about 6% as large as the effect of a 100-point increase in the focal student's prior GPA, and it is about 10% of the student's own GPA. We reviewed the literature and found that prior studies reported effect sizes at a similar scale for various environments (e.g., dormitories and classrooms) and cultures (e.g., Western universities). For instance, Sacerdote (2001) found that a 1-*s.d.* increase in the roommate's GPA is associated with a 0.05 increase in the student's freshmen year GPA (an effect size of about 4% of their own GPA)[12], Zimmerman (2003) found that a 100-point increase in roommate verbal SAT score translated into a 0.03 increase in the student's first-semester GPA (an effect size of about 15% as large as a 100-point increase in their own verbal SAT score)[24], Carman et al. (2012) found that a 0.1-*s.d.* increase in peer average math achievement increases the student's math test score by 0.04 *s.d.* on average[9], and Thiemann (2021) reported negative results that 1-*s.d.* increase in average peer ability is associated with a decrease in a student's GPA by 7.9% of 1-*s.d.* on average[45]. Presumedly, other external factors may contribute to this observation, such as that dorm rooms are not formal learning environments like classrooms and libraries, and there are other channels through which students can learn from and influence each other.

**5.2 Regression on the effects of in-dorm ordinal rank**

Competitive dynamics between roommates in the interpersonal and local dorm room environment could also affect student academic performance[46-48]. For example, on one extreme, in a dorm room that contains four students all with high academic performance (e.g., GPA >= 0.9), there is a possibility that one student's ordinal rank in the dorm room is always 4th across different semesters (e.g., GPA = 0.9, which is smaller than the GPAs of their three roommates). In this case, even if the student has high academic performance, they may still feel discouraged and less motivated due to their consistently lowest relative academic performance in the dorm room. On the other extreme, one student's ordinal rank in the dorm room is always 1st, conditional on absolute academic performance, they may feel more confident than others and have better long-term outcomes[49-51]. This prompts us to study how a student's post-GPA is impacted by their in-dorm ordinal rank according to prior GPA (OR_InDorm), which is the number of better-achieving roommates including themselves, with 1 being the highest and 4 being the lowest in performance.

Specifically, in the analysis, we employ an OLS model to study how a student's in-dorm ordinal rank (OR_InDorm) affects their future performance (GPA_Post) after controlling their prior performance, the average and differences in roommate prior performance, their gender, major, cohort, and semester. Meanwhile, we examine how peer heterogeneity in prior performance moderates this effect. Specifically, the empirical specification is given by

$$G_i^{s+1} = b_0 + b_1 G_i^s + b_2 OR_i^s + b_3 RA_i^s + b_4 RD_i^s + b_5 OR_i^s \times G_i^s + b_6 OR_i^s \times RA_i^s$$
$$+ b_7 OR_i^s \times RD_i^s + b_8 D^{Ge} + b_9 D^{Ma} + b_{10} D^{Co} + b_{11} D^{Se} + \epsilon_i, \qquad (9)$$

where $OR_i^s$ is the OR_InDorm of student $i$ in semester $s$ (ranging from 1 to 4) and $\epsilon_i$ is the error term. The interaction terms are $OR_i^s \times G_i^s$ between OR_InDorm and GPA_Prior, $OR_i^s \times RA_i^s$ between OR_InDorm and RM_Avg, and $OR_i^s \times RD_i^s$ between OR_InDorm and RM_Diff for student $i$ in semester $s$. All other controls are the same as introduced above.

The regression results for all five semesters are summarized in Table 2 of the main text. We find that (1) in-dorm ordinal rank has positive effects on post-GPA, (2) neither the interaction term of in-dorm ordinal rank and prior GPA nor the interaction term of in-dorm ordinal rank and roommate average prior GPA is significant, and (3) the interaction term of in-dorm ordinal rank and the differences in roommate prior GPA is significantly negative. The results suggest the significant moderating role of peer heterogeneity. In Fig. 4b of the main text, we show the margin plot of the linear relationship between GPA_Post and OR_InDorm, which is moderated by RM_Diff. To assess the significance of the moderating effect of RM_Diff on the relationship, we perform a simple slope testing and summarize the results in Supplementary Figure 11b. We find that the slope decreases with the increase of RM_Diff and becomes not statistically different from 0 when RM_Diff > −0.3 *s.d.* (*P* value > 0.1). In particular, the slope $b = 0.007$ (*s.e.* = 0.003; 95% CI = [0.002, 0.012]; *P* value = 0.009) when RM_Diff is low (i.e., 1 *s.d.* below its mean), and the slope $b = -0.000$ (*s.e.* = 0.004; 95% CI = [−0.007, 0.007]; *P* value = 0.982) when RM_Diff is high (i.e., 1 *s.d.* above its mean). These results suggest that the positive relationship between a student's post-GPA and their roommate average prior GPA is more pronounced when the differences in their roommate prior GPA are low but loses significance when the differences are high.

## 5.3 Falsification tests of the regression results

To check the robustness of the OLS regression results and demonstrate significance, we conduct a falsification test by running the same OLS regressions on the roommate null model that consists of 100 independent implementations (see Section 3.1 for details). In the null model, students are assumed to live together for all semesters once assigned to the same room. The falsification testing results are summarized in Supplementary Figure 12 and Supplementary Figure 13.

Analyzing the OLS regression results based on the roommate null model, we find that the effect of roommate average prior performance (RM_Avg) on a student's future performance (GPA_Post) is very small ($b = 0.004$ in the null model vs $0.050$ in the actual data, both with $P$ value $< 0.01$), and the moderation effect of the differences in roommate prior performance (RM_Diff) is also very small ($b = –0.011$ in the null model vs $–0.089$ in the actual data, both with $P$ value $< 0.1$; see Supplementary Figure 12 and Fig. 3 of the main text). Although simple slope testing results show that, in the roommate null model, the effect of RM_Avg on GPA_Post is more pronounced and significant ($b = 0.004$, $P$ value $< 0.01$) when RM_Diff is low (i.e., 1 $s.d.$ below its mean; Supplementary Figure 14a), the effect is still much smaller than that in the actual data ($b = 0.066$, $P$ value $< 0.01$; Supplementary Figure 11a). Together, these results show that the effect size of RM_Avg and RM_Diff on GPA_Post in the roommate null model is only about 10% of that in the actual data, which suggests the significance of the findings from the regression analysis.

Moreover, we find that the effect of a student's within-dorm ordinal rank according to prior performance (OR_InDorm) on the student's future performance (GPA_Post) is very small ($b = 0.001$ in the null model vs $0.006$ in the actual data; both with $P$ value $< 0.05$), and the moderating effect of RM_Diff is also very small ($b = –0.007$ in the null model with $P$ value $< 0.01$ vs $–0.022$ in the actual data with $P$ value $< 0.05$; see Supplementary Figure 13 and Fig. 4 of the main text). These results show that the effect size of OR_InDorm and RM_Diff on GPA_Post in the null model is about 20% of that in the actual data. Interestingly, we notice that the relationship between OR_InDorm and GPA_Post is significantly negative in the null model ($b = –0.0013$, $P$ value $< 0.01$) when RM_Diff is high (i.e., 1 $s.d.$ above its mean; see Supplementary Figure 13c and Supplementary Figure 14b). This is because when RM_Diff is high, the student can only have a high GPA_Prior when their OR_InDorm is low (e.g., ranked 1$^{st}$ in the dorm room) and a low GPA_Prior when their OR_InDorm is high (e.g., ranked 4$^{th}$ in the dorm room). There, the high correlation between GPA_Prior and GPA_Post leads to a moderate negative relationship between OR_InDorm and GPA_Post. The significantly positive relationship between OR_InDorm and GPA_Post in the null model ($b = 0.0008$, $P$ value $< 0.01$) when RM_Diff is low (i.e., 1 $s.d.$ below its mean) is, however, due to "regression toward the mean" when all dorm rooms are included in the regression (Supplementary Figure 13c and Supplementary Figure 14b).

Taken together, these results from the falsification tests on the roommate null model support our findings that roommate peer effects on academic performance are nontrivial and significant although the magnitude of the effects appears to be not large (see Section 5.1 for details).

# Supplementary Figures

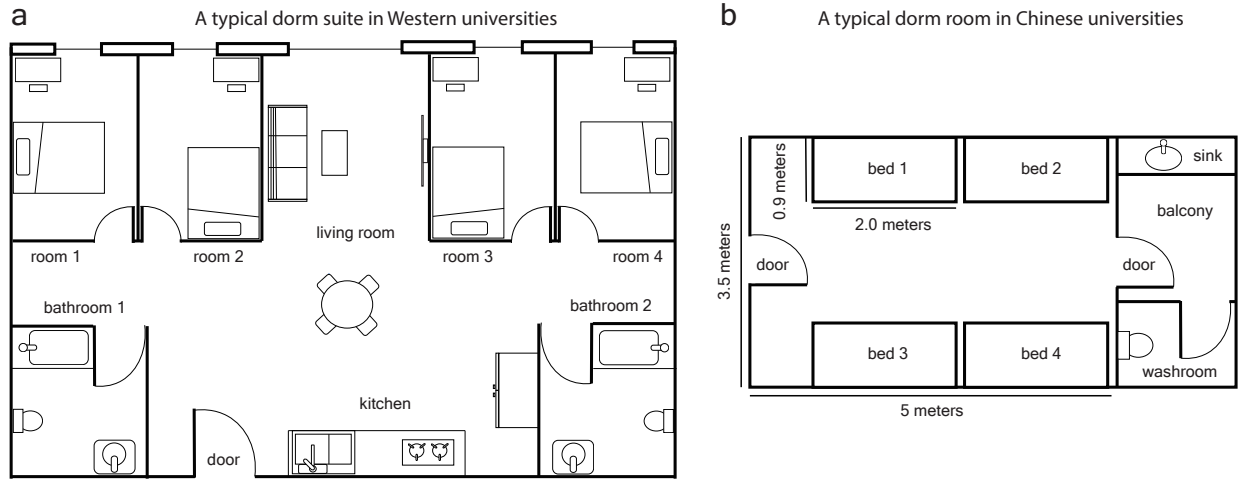

**Supplementary Figure 1. Examples of dorm suites in Western universities and dorm rooms in Chinese universities.** (a) Floor plan of a typical dorm suite in Western universities. A 4-person dorm suite has four separate bedrooms, where each student occupies one bedroom and shares public areas with suitemates such as a living room, kitchen, and bathroom. The dorm suite offers a private and independent space for each student. (b) Floor plan of a typical dorm room in Chinese universities. A 4-person dorm room contains four beds, where each student occupies one bed and shares public areas with roommates such as a balcony and washroom. The dorm room provides a small but interpersonal space for all roommates to frequently interact with each other.

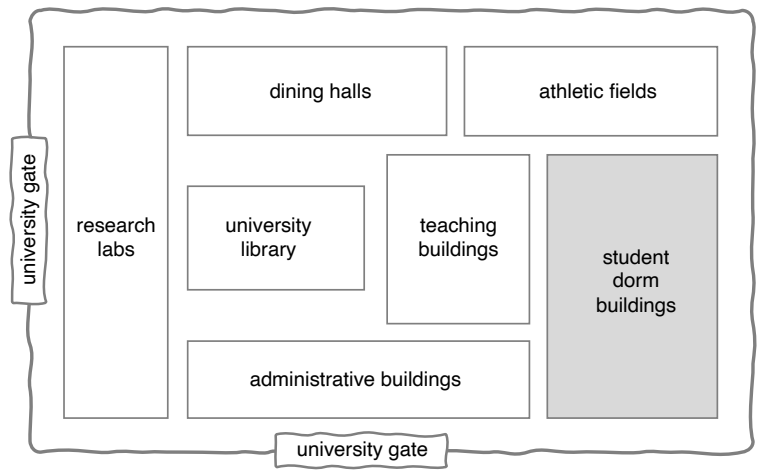

**Supplementary Figure 2. A sketch of the campus facility layout of the university under study.** The university campus has clear geographical boundaries protected by fences, and gates are guarded for security reasons. Dorm buildings are all within the university's geographic boundary (i.e., on campus), and they are identical in physical environment and very close to each other in distance.

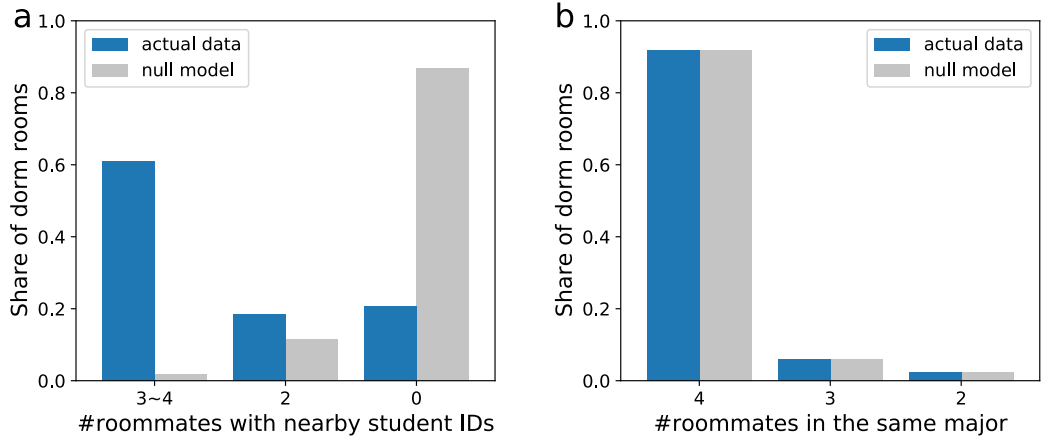

**Supplementary Figure 3. Some patterns of the initial roommate assignment.** (a) Roommates tend to have nearby student IDs. In more than 60% of dorm rooms, at least three roommates have nearby student IDs in the actual data. By comparison, this fraction is less than 2% in the roommate null model. (b) Roommates tend to have the same major. In more than 90% of dorm rooms, four roommates have the same major. The results are the same in the actual data and the roommate null model because the null model controls for the student's major in the random shuffling process.

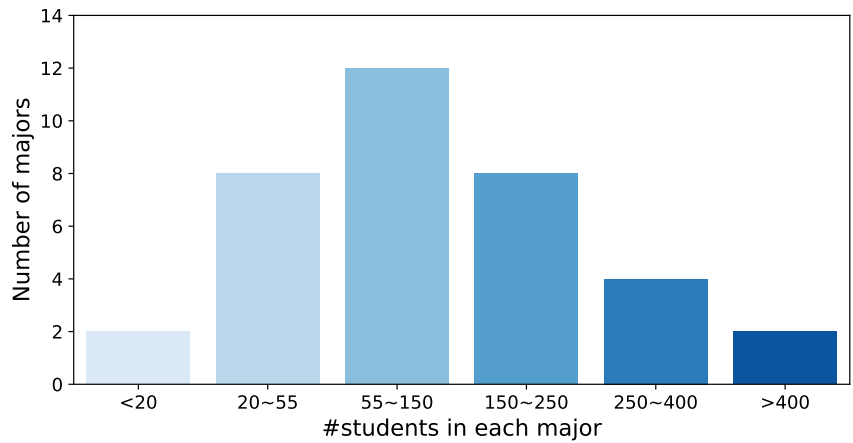

**Supplementary Figure 4. Student compositions in each major or dorm room.** Distribution of majors by the number of students per major. Most majors have 55 to 150 students. One major has fifteen students at the minimum and as many as over 400 students in the sample.

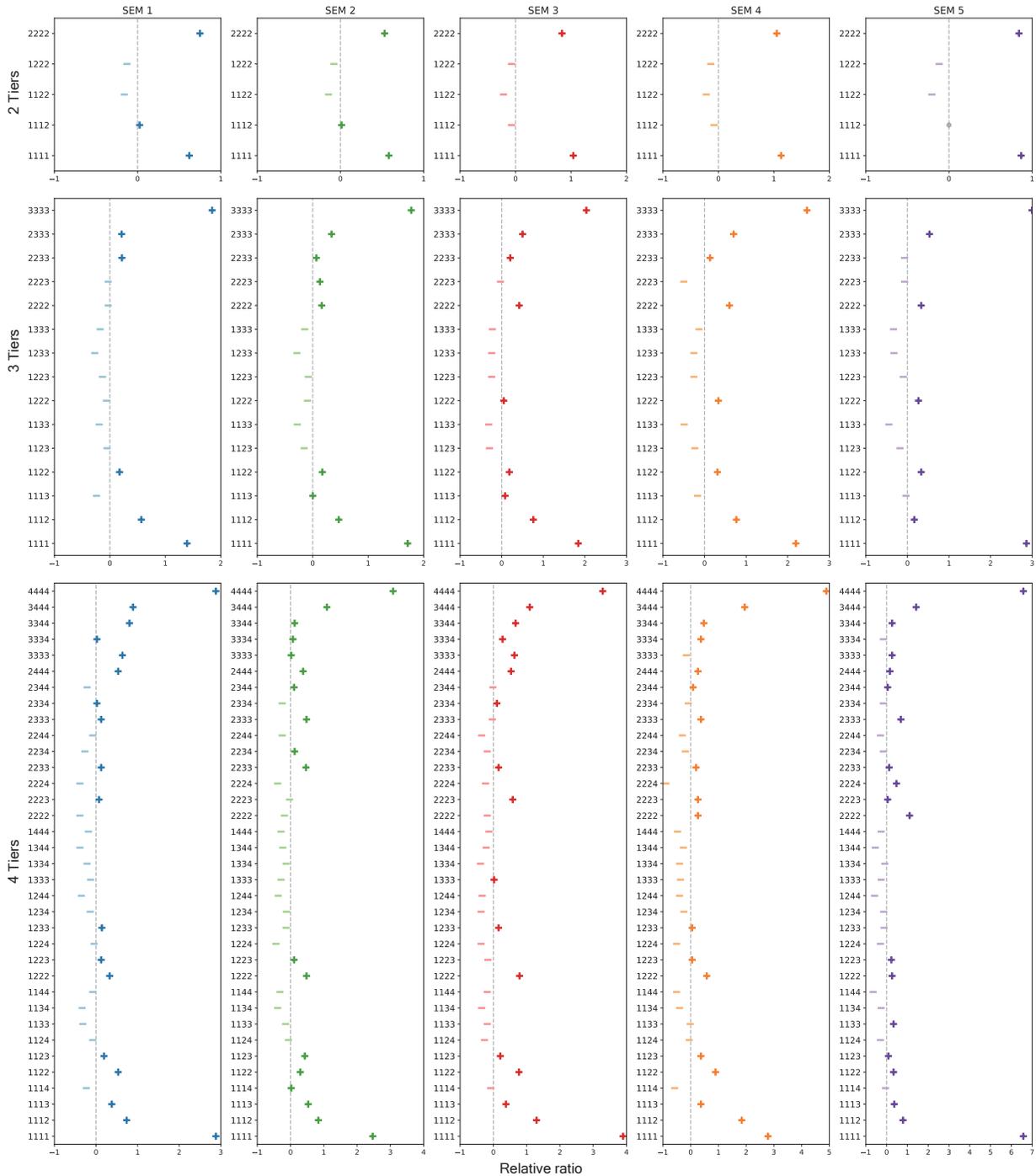

**Supplementary Figure 5. The relative ratio $\mathbb{E}$ of each tier combination in each semester.** Each row shows the $\mathbb{E}$ of each combination under the 2-tier, 3-tier, and 4-tier classification of GPA, respectively. Each column corresponds to one semester, from semester 1 to semester 5. The vertical axis shows all unique combinations in ascending order of tier numbers under a given tier of classification, and the horizontal axis shows $\mathbb{E}$ which compares the actual and theoretical probability of the same combination. The vertical dashed line marks $E = 0$, where the actual and theoretical probabilities are equal. Positive and negative $E$ are marked by '+' and '-', respectively.

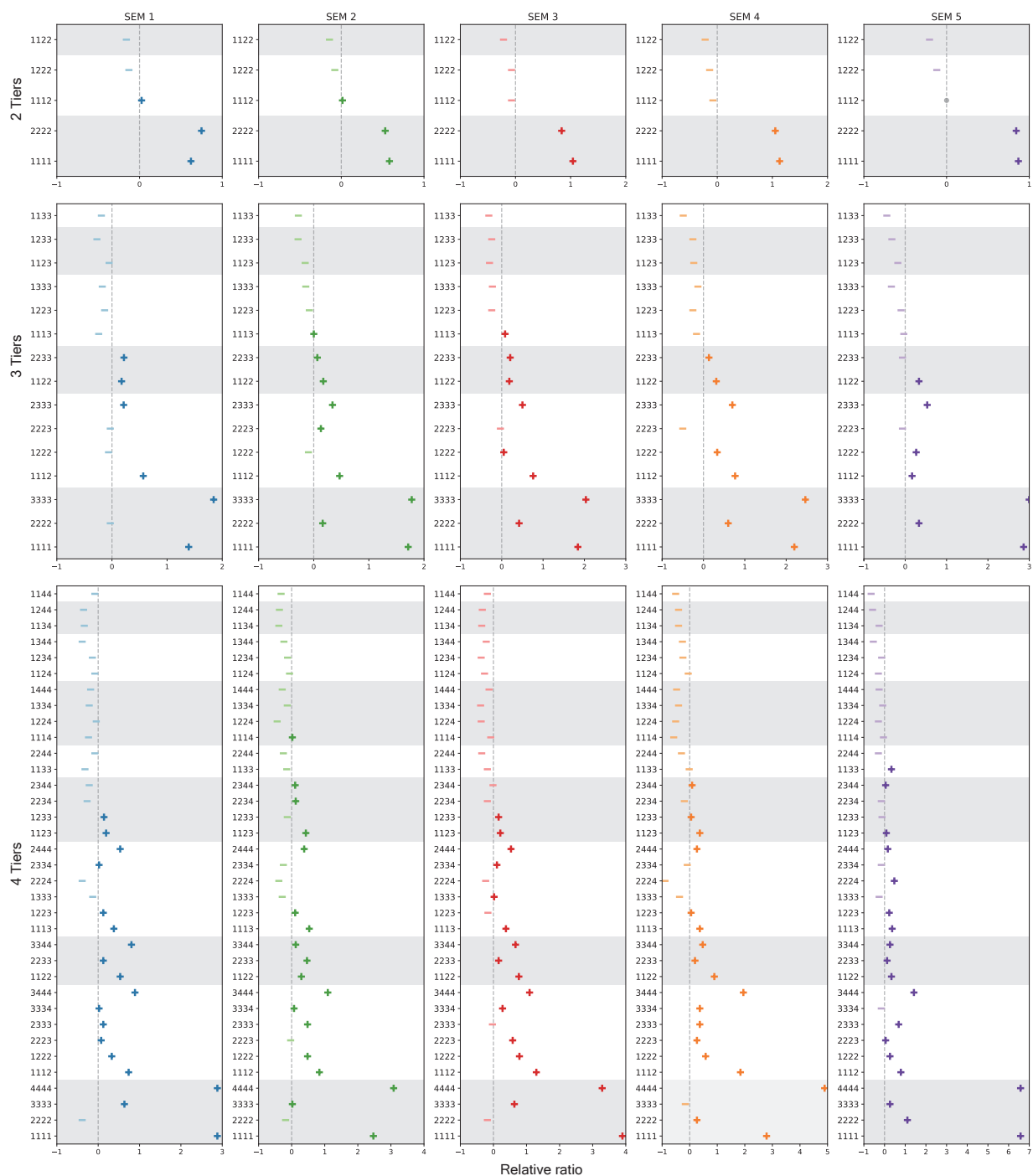

**Supplementary Figure 6. Rearranged tier combinations according to their relative difference.** The vertical axis shows combinations in ascending order of the relative difference $D$, which measures the average pairwise difference among tier numbers. The staggered gray shade marks a group of combinations with the same $D$. The horizontal axis shows $\mathbb{E}$ which compares the actual probability with the theoretical probability of the same combination. Other captions are the same as those for Supplementary Figure 5.

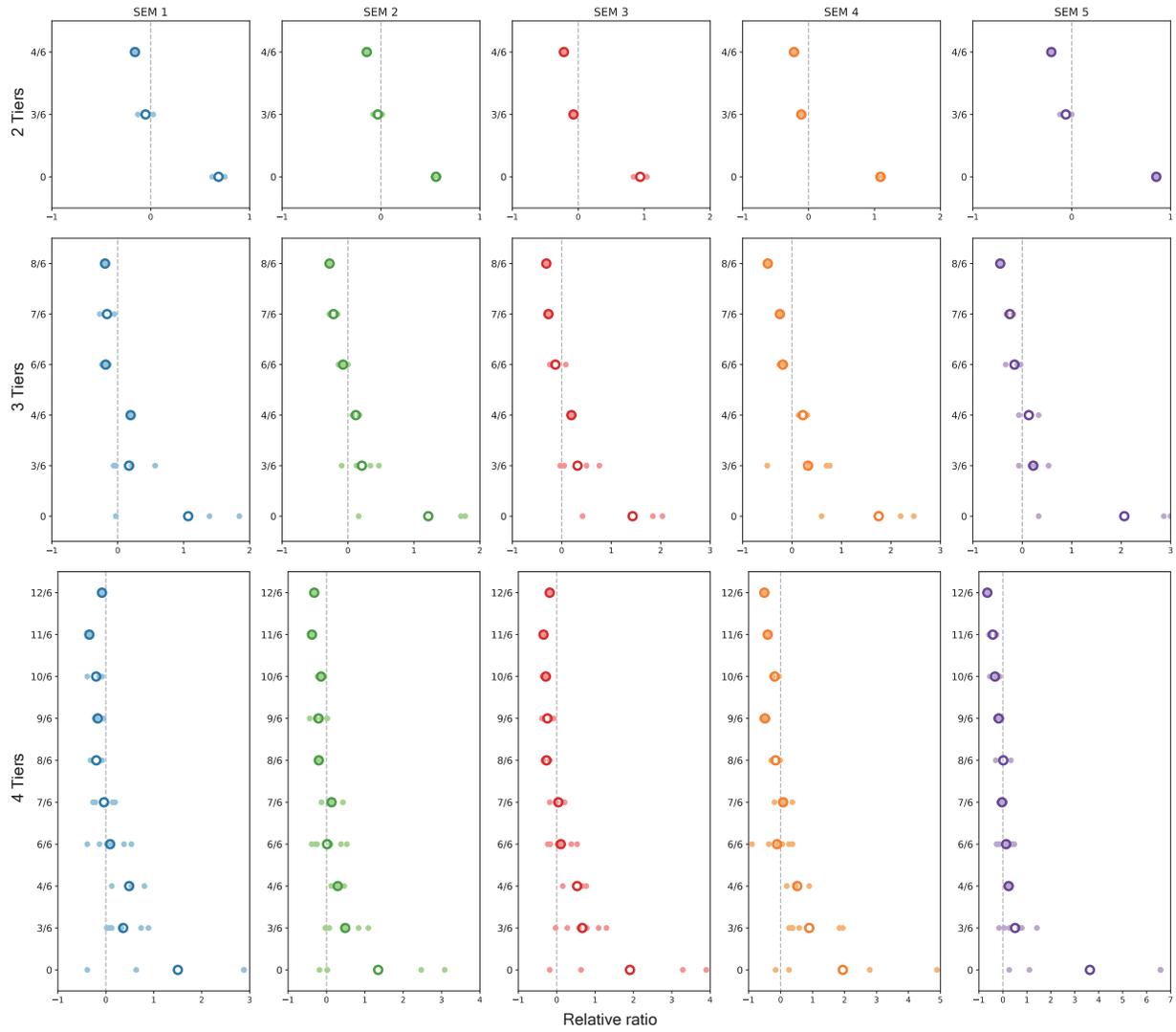

**Supplementary Figure 7. The negative association between the relative ratio $\mathbb{E}$ and the relative difference $D$.** The horizontal axis shows the relative ratio $\mathbb{E}$, and the vertical axis shows the relative difference $D$. Combinations with the same relative difference $D$ are grouped and their relative ratios $\mathbb{E}$ are averaged. The hollow circle shows the mean $\mathbb{E}$ for each group with the same $D$, and data points show $\mathbb{E}$ for each combination. Other captions are the same as those for Supplementary Figure 5.

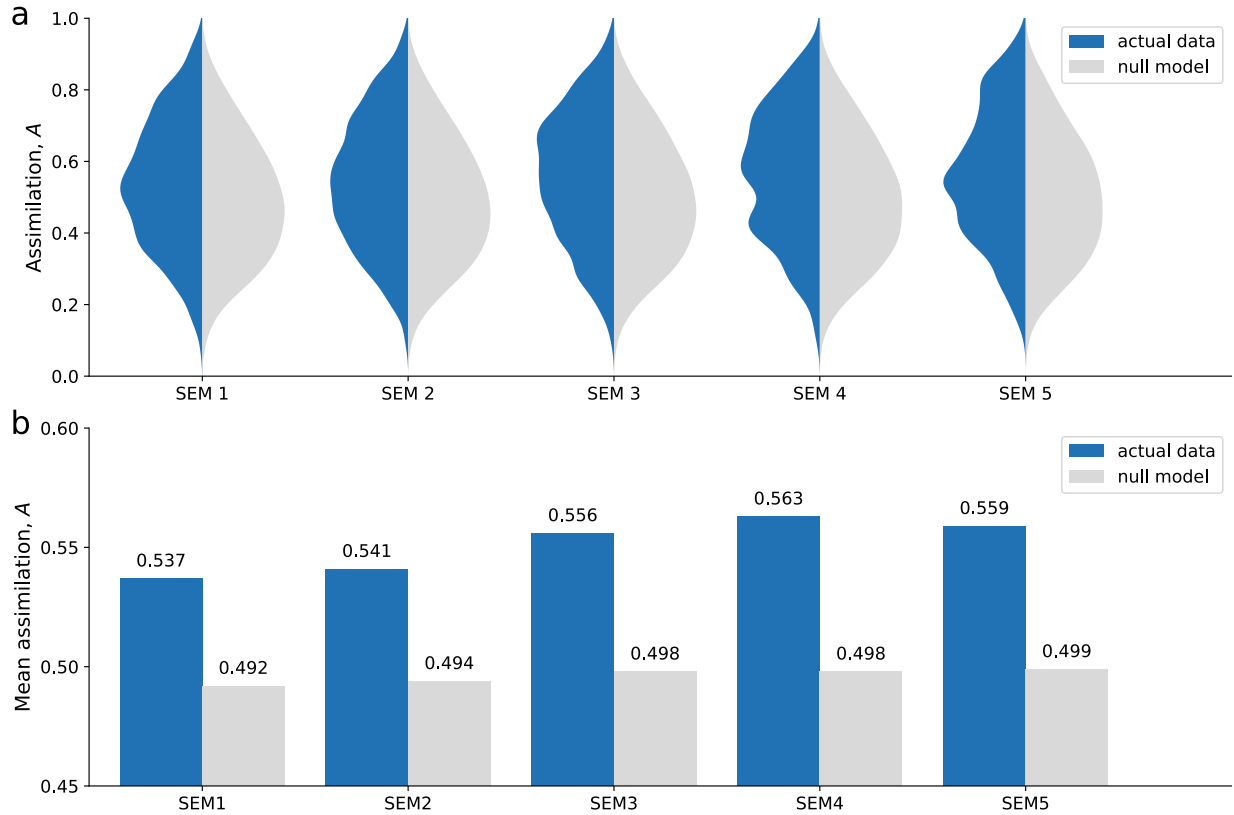

**Supplementary Figure 8. The assimilation of roommate academic performance in each semester.** (a) The density distribution $p(A)$ of assimilation for all dorm rooms. In the distribution plot, the left half (in blue) shows the actual assimilation and the right half (in gray) shows the null-model assimilation. (b) The comparison between the mean actual assimilation and the mean null-model assimilation from semester 1 to semester 5. The mean value is labeled on each bar.

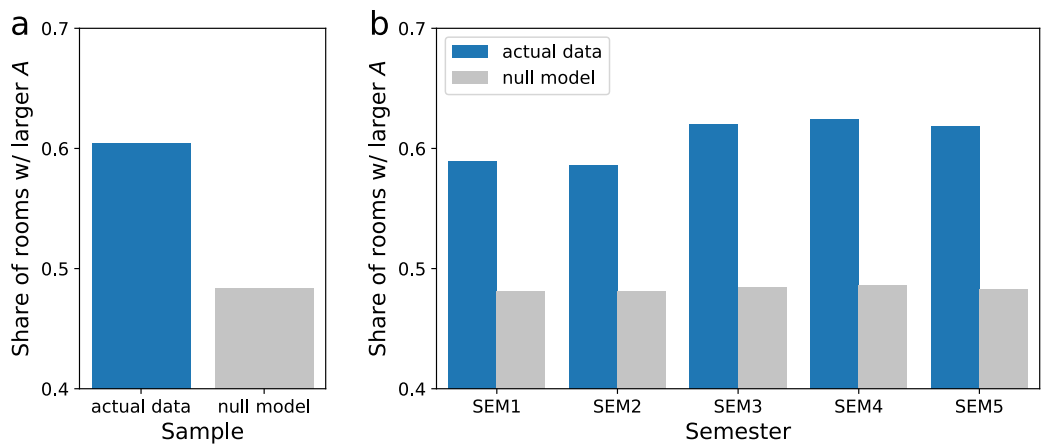

**Supplementary Figure 9. The assimilation of roommate academic performance and its temporal trend.** (a) The share of dorm rooms with larger-than-null-model assimilation according to the 95% confidence interval. (b) The results for each of the five semesters.

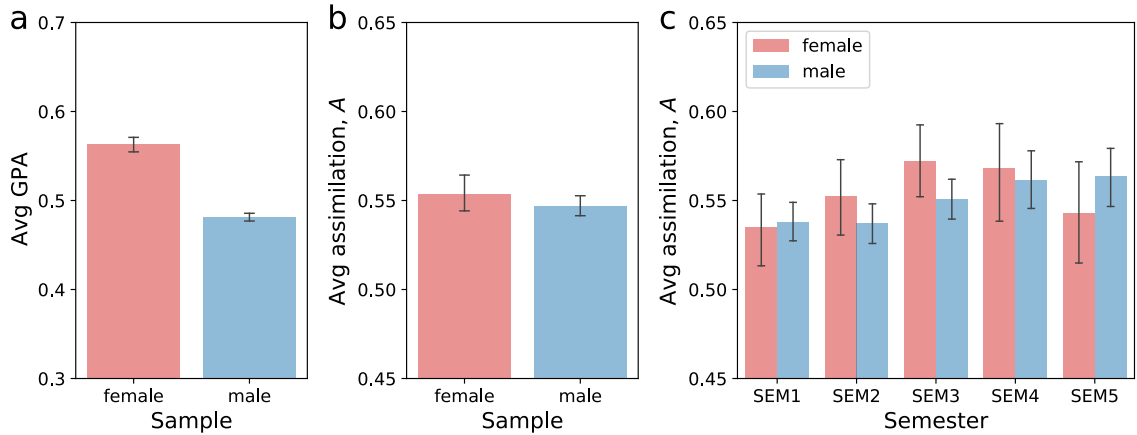

**Supplementary Figure 10. The roommate peer effects for female and male students.** (a) The average GPA for all five semesters. Error bars represent standard errors clustered for students. (b) The average assimilation for all five semesters. (c) The average assimilation for each of the five semesters. Error bars represent standard errors clustered for dorm rooms.

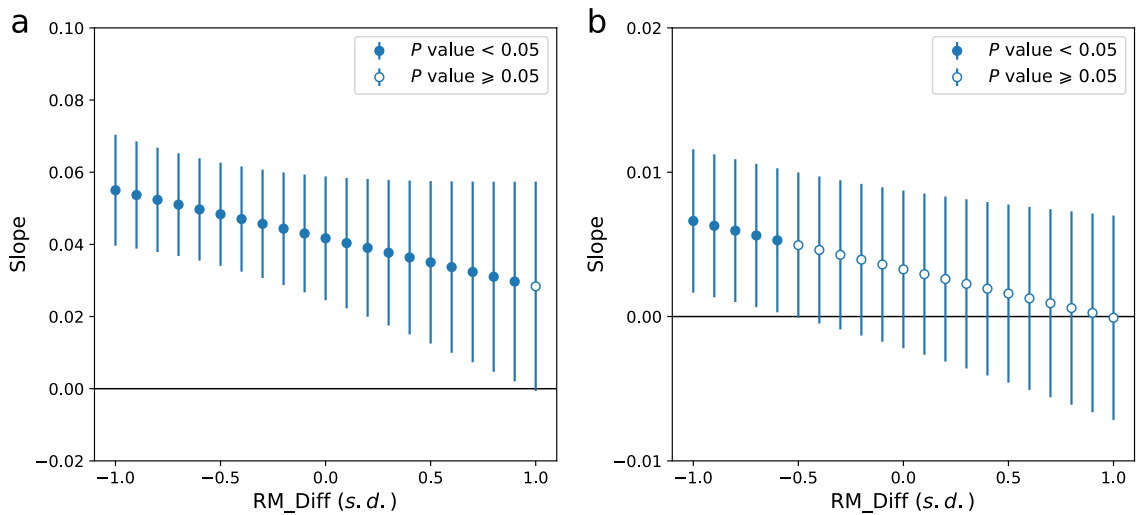

**Supplementary Figure 11. Results of the simple slope tests on moderation effects.** (a) The slope for the relationship between a student's post-GPA and their roommate average prior GPA. The relationship is moderated by the differences in roommate prior GPAs (RM_Diff), and the slope is significantly above 0 ($P$ value < 0.1). (b) The slope for the relationship between a student's post-GPA and their in-dorm ordinal rank (OR_InDorm). The relationship is moderated by RM_Diff, but only when RM_Diff < −0.5 $s.d.$, the slope is significantly above 0 ($P$ value < 0.05). In both panels, the x-axis shows the differences ranging from −1 $s.d.$ to 1 $s.d.$ of RM_Diff, and the y-axis shows the slope of the corresponding relationship. Solid circles show a $P$ value < 0.05, and hollow circles show a $P$ value ≥ 0.05. Error bars represent the 95% confidence intervals.

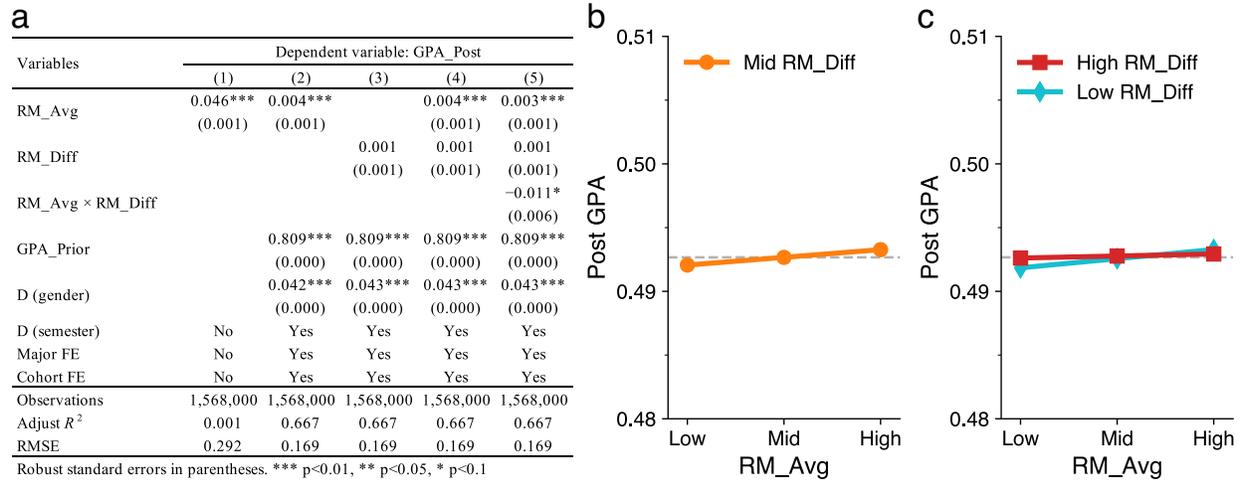

**Supplementary Figure 12. Results of the falsification test for the effects of roommate average prior performance.** The OLS regression is applied to the roommate null model that consists of 100 independent implementations. (a) Full regression results. (b) The margin plot for the linear relationship between GPA_Post and RM_Avg. (c) The margin plot for the moderating effects of RM_Diff on the relationship between GPA_Post and RM_Avg. The "Low" and "High" on the horizontal axis represent 1 standard deviation (*s.d.*) below and above the mean ("Mid") of RM_Avg, respectively. The "Low" and "High" in the legend represent 1 *s.d.* below and above the mean ("Mid") of RM_Diff, respectively. The horizontal dashed line marks the regression constant.

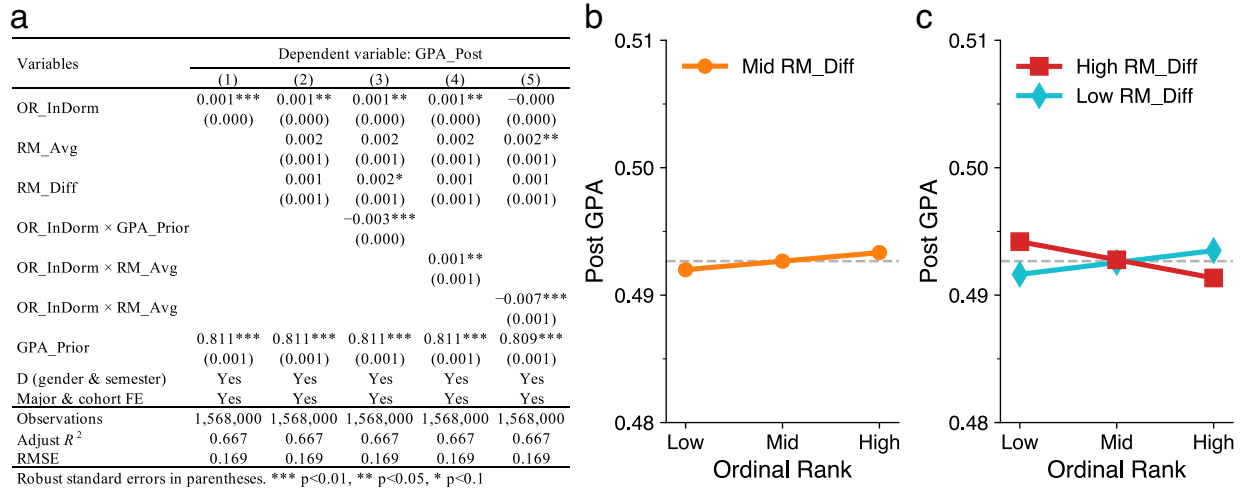

**Supplementary Figure 13. Results of the falsification test for the effects of within-dorm ordinal rank.** The OLS regression is applied to the roommate null model that consists of 100 independent implementations. (a) Full regression results. (b) The margin plot for the linear relationship between GPA_Post and OR_InDorm, where a larger value means a relatively low GPA among roommates. (c) The margin plot for the moderating effect of RM_Diff on the relationship between GPA_Post and OR_InDorm. The "Low" and "High" on the horizontal axis represent 1 standard deviation (*s.d.*) below and above the mean ("Mid") of OR_InDorm, respectively. The "Low" and "High" in the legend represent 1 *s.d.* below and above the mean ("Mid") of RM_Diff, respectively. Horizontal dashed lines mark regression constants.

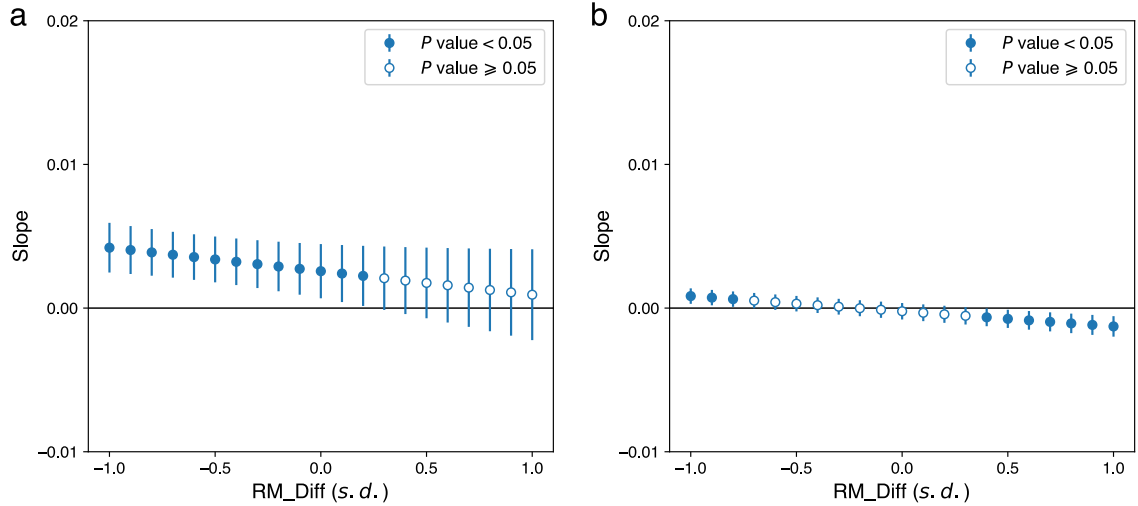

**Supplementary Figure 14. Results of the simple slope tests on moderation effects.** (a) The slope for the relationship between a student's post-GPA and their roommate average prior GPA. The relationship is moderated by the differences in roommate prior GPAs (RM_Diff), and the slope is significantly above 0 only when RM_Diff < 0.3 *s.d.* ($P$ value < 0.05). (b) The slope for the relationship between a student's post-GPA and their in-dorm ordinal rank (OR_InDorm). The relationship is moderated by RM_Diff. The slope is significantly above 0 when RM_Diff < −0.7 *s.d.* and significantly below 0 when RM_Diff > 0.3 *s.d.* ($P$ value < 0.05). In both panels, the x-axis shows the differences ranging from −1 *s.d.* to 1 *s.d.* of RM_Diff, and the y-axis shows the slope of the corresponding relationship. Solid circles show a $P$ value < 0.05, and hollow circles show a $P$ value ≥ 0.05. Error bars represent the 95% confidence intervals.

# Supplementary Tables

**Supplementary Table 1. The theoretical probability of each tier combination.**

| Tier | 1111 | 1112 | 1113 | 1114 | 1122 | 1123 | 1124 | 1133 | 1134 | 1144 | 1222 | 1223 |
|---|---|---|---|---|---|---|---|---|---|---|---|---|
| 2 | 1/16 | 1/4 | – | – | 3/8 | – | – | – | – | – | 1/4 | – |
| 3 | 1/81 | 4/81 | 4/81 | – | 2/27 | 4/27 | – | 2/27 | – | – | 4/81 | 4/27 |
| 4 | 1/256 | 1/64 | 1/64 | 1/64 | 3/128 | 3/64 | 4/64 | 3/128 | 3/64 | 3/128 | 1/64 | 3/64 |

| Tier | 1224 | 1233 | 1234 | 1244 | 1333 | 1334 | 1344 | 1444 | 2222 | 2223 | 2224 | 2233 |
|---|---|---|---|---|---|---|---|---|---|---|---|---|
| 2 | – | – | – | – | – | – | – | – | 1/16 | – | – | – |
| 3 | – | 4/27 | – | – | 4/81 | – | – | – | 1/81 | 4/81 | – | 2/27 |
| 4 | 3/64 | 3/64 | 3/32 | 3/64 | 1/64 | 3/64 | 3/64 | 1/64 | 1/256 | 1/64 | 1/64 | 3/128 |

| Tier | 2234 | 2244 | 2333 | 2334 | 2344 | 2444 | 3333 | 3334 | 3344 | 3444 | 4444 |
|---|---|---|---|---|---|---|---|---|---|---|---|
| 2 | – | – | – | – | – | – | – | – | – | – | – |
| 3 | – | – | 4/81 | – | – | – | 1/81 | – | – | – | – |
| 4 | 3/64 | 3/128 | 1/64 | 3/64 | 3/64 | 1/64 | 1/256 | 1/64 | 3/128 | 1/64 | 1/256 |

*Notes:* The total number of combinations are 5, 15, and 35 when applying 2-tier, 3-tier, and 4-tier classifications, respectively. A combination has no particular tier order, so the one in ascending tier order is used to delegate all identical others. Each title row shows combinations, and the value in the table is the theoretical probability of each combination, respectively. The symbol "–" means that there is no such combination under the corresponding tier classification.

**Supplementary Table 2. The significance of each tier combination in the actual data.**

| Tier | 1111 | 1112 | 1113 | 1114 | 1122 | 1123 | 1124 | 1133 | 1134 | 1144 | 1222 | 1223 |
|---|---|---|---|---|---|---|---|---|---|---|---|---|
| 2 | *** | (***) | – | – | (***) | – | – | – | – | – | (**) | – |
| 3 | *** | *** | (**) | – | ** | (***) | – | (***) | – | – | n.s. | (***) |
| 4 | *** | *** | *** | (*) | *** | n.s. | (***) | (**) | (***) | (***) | * | n.s. |

| Tier | 1224 | 1233 | 1234 | 1244 | 1333 | 1334 | 1344 | 1444 | 2222 | 2223 | 2224 | 2233 |
|---|---|---|---|---|---|---|---|---|---|---|---|---|
| 2 | – | – | – | – | – | – | – | – | *** | – | – | – |
| 3 | – | (***) | – | – | n.s. | – | – | – | *** | n.s. | – | *** |
| 4 | (***) | (**) | (***) | (***) | (*) | (***) | (**) | n.s. | * | *** | n.s. | *** |

| Tier | 2234 | 2244 | 2333 | 2334 | 2344 | 2444 | 3333 | 3334 | 3344 | 3444 | 4444 |
|---|---|---|---|---|---|---|---|---|---|---|---|
| 2 | – | – | – | – | – | – | – | – | – | – | – |
| 3 | – | – | *** | – | – | – | *** | – | – | – | – |
| 4 | n.s. | (*) | ** | n.s. | ** | *** | n.s. | *** | *** | *** | *** |

*Notes:* A statistical test is performed using the roommate null model to examine the significance of each combination's relative difference $\mathbb{E}$ under the 2-tier, 3-tier, and 4-tier classifications, respectively (see Section 3.2). The null model consists of 1000 independent implementations, and the statistical test is one-sided. Each title row shows combinations. The value in the table marks the significance of the combination's $\mathbb{E}$, where *, **, and *** represent significantly positive $\mathbb{E}$ with a $P$ value < 0.1, 0.05, and 0.01, respectively. The symbol "(·)" means significantly negative $\mathbb{E}$ at the corresponding level, "n.s." means not significantly different from 0 (i.e., $P$ value > 0.1), and "–" means that there is no such combination under the corresponding tier classification.

**Supplementary Table 3. Summary statistics of variables and their correlations.**

| No. | Variable | Symbol | Mean | Std. Dev. | Pearson's correlation coefficient ||||||||
|---|---|---|---|---|---|---|---|---|---|---|---|---|
| | | | | | 1 | 2 | 3 | 4 | 5 | 6 | 7 | 8 |
| 1 | GPA_Post | $G^{s+1}$ | 0.500 | 0.292 | | | | | | | | |
| 2 | GPA_Prior | $G^s$ | 0.500 | 0.292 | 0.815 | | | | | | | |
| 3 | RM_Avg | $RA^s$ | 0.500 | 0.195 | 0.244 | 0.251 | | | | | | |
| 4 | RM_Diff | $RD^s$ | 0.303 | 0.150 | 0.020 | 0.020 | 0.024 | | | | | |
| 5 | OR_InDorm | $OR^s$ | 2.500 | 1.118 | −0.539 | −0.694 | 0.347 | −0.000 | | | | |
| 6 | D (gender) | $D^{Ge}$ | 0.226 | 0.418 | 0.132 | 0.108 | 0.162 | −0.009 | 0.000 | | | |
| 7 | D (major) | $D^{Ma}$ | 10.750 | 8.562 | 0.000 | 0.000 | −0.007 | 0.045 | −0.005 | 0.088 | | |
| 8 | D (cohort) | $D^{Co}$ | 1.345 | 0.475 | 0.000 | −0.000 | 0.000 | 0.018 | 0.000 | 0.019 | −0.013 | |
| 9 | D (semester) | $D^{Se}$ | 2.155 | 1.064 | 0.000 | −0.000 | −0.000 | −0.043 | −0.000 | −0.009 | 0.006 | −0.447 |

*Notes:* GPA_Post and GPA_Prior are the student's GPA in semester $s+1$ and $s$, respectively. The semester index $s$ ranges from 1 to 4. RM_Avg is the average roommate prior GPA. RM_Diff is the differences in roommate prior GPAs. OR_InDorm is the student's in-dorm ordinal rank with 1 being the highest and 4 being the lowest in ranking (i.e., from the highest to the lowest academic performance). The gender dummy D (gender) is coded as 1 and 0 for females and males, respectively. The major dummy D (major) is coded from 1 to 36 from the largest to the smallest major according to the number of students. The cohort dummy D (cohort) is coded as 1 and 2 for the 2011 and 2012 cohort, respectively. The semester dummy D (semester) is coded from 1 to 4 from the first to the fourth semester. Data from all five semesters are pooled together.